\documentclass[10pt, journal]{IEEEtran}

\usepackage{amsthm} 
\usepackage{amsmath}    
\IEEEoverridecommandlockouts
\usepackage{empheq}
\usepackage{bm}
\usepackage{soul}
\usepackage{graphicx}
\usepackage{indentfirst}
\usepackage{epstopdf}
\usepackage{amssymb}
\usepackage{url}
\usepackage{enumitem} 
\usepackage{multirow}
\usepackage{hhline}
\usepackage{booktabs}
\usepackage{mathtools}
\usepackage{makecell}
\usepackage[linesnumbered,boxed,commentsnumbered,ruled,vlined,longend]{algorithm2e}
\usepackage{comment}

\DeclareMathOperator*{\minimize}{minimize}

\DeclareMathOperator*{\subjectto}{subject\ to}

\makeatother
\DeclareMathAlphabet\mathbfcal{OMS}{cmsy}{b}{n}

\newtheorem{theorem}{Theorem}
\newtheorem{mydef}{Definition}

\newtheorem{asmp}{Assumption}



\usepackage{stackengine}

\newcommand{\mat}[1]{\boldsymbol{#1}}

\newcommand{\bmat}[1]{\begin{bmatrix} #1 \end{bmatrix}}

\providecommand{\mA}{\ensuremath{\mat{A}}}
\providecommand{\mB}{\ensuremath{\mat{B}}}
\providecommand{\mC}{\ensuremath{\mat{C}}}
\providecommand{\mD}{\ensuremath{\mat{D}}}
\providecommand{\mE}{\ensuremath{\mat{E}}}
\providecommand{\mF}{\ensuremath{\mat{F}}}
\providecommand{\mG}{\ensuremath{\mat{G}}}
\providecommand{\mH}{\ensuremath{\mat{H}}}
\providecommand{\mI}{\ensuremath{\mat{I}}}

\providecommand{\mK}{\ensuremath{\mat{K}}}

\providecommand{\mN}{\ensuremath{\mat{N}}}
\providecommand{\mO}{\ensuremath{\mat{O}}}
\providecommand{\mP}{\ensuremath{\mat{P}}}
\providecommand{\mQ}{\ensuremath{\mat{Q}}}
\providecommand{\mR}{\ensuremath{\mat{R}}}
\providecommand{\mS}{\ensuremath{\mat{S}}}

\providecommand{\mW}{\ensuremath{\mat{W}}}
\providecommand{\mX}{\ensuremath{\mat{X}}}

\newcommand{\m}{\boldsymbol}
\allowdisplaybreaks[4]
\usepackage[colorlinks = true,
linkcolor = blue,
urlcolor  = blue,
citecolor = blue,
anchorcolor = blue]{hyperref}

\newcommand{\mc}[1]{\mathcal{#1}}
\newcommand{\mbb}[1]{\mathbb{#1}}
\newcommand{\mr}[1]{\mathrm{#1}}
\usepackage[framemethod=TikZ]{mdframed}
\mdfdefinestyle{MyFrame}{%
	linecolor=black,
	outerlinewidth=1.25pt,
	roundcorner=1.25pt,
	innerrightmargin=5pt,
	innerleftmargin=5pt,}
	
\usepackage[noabbrev]{cleveref}

\usepackage{mathtools}

\DeclarePairedDelimiter\abs{\lvert}{\rvert}%
\DeclarePairedDelimiter\norm{\lVert}{\rVert}%

\makeatletter
\let\oldabs\abs
\def\abs{\@ifstar{\oldabs}{\oldabs*}}
\let\oldnorm\norm
\def\norm{\@ifstar{\oldnorm}{\oldnorm*}}
\makeatother

\usepackage[english]{babel}
\usepackage[utf8]{inputenc}
\usepackage[super]{nth}

\usepackage{float}
\usepackage[caption = false]{subfig}

\usepackage{array}
\usepackage{threeparttable}

\renewcommand\IEEEkeywordsname{Keywords}

\usepackage[english]{babel}
\usepackage[utf8]{inputenc}
\usepackage[super]{nth}

\RequirePackage{filecontents}

\SetKwRepeat{Do}{do}{while}%

\usepackage{lipsum}
\usepackage{cite}

\usepackage{lettrine}

\usepackage{eqnarray,amsmath}

\title{\centering \Huge {{Robust Feedback Control of Power Systems with Solar Plants and Composite Loads}}}
\author{Muhammad Nadeem,  MirSaleh Bahavarnia, and Ahmad F. Taha \vspace{-0.45cm}
	\thanks{The authors are with the Civil and Environmental Engineering Department, Vanderbilt University, 2201 West End Ave, Nashville, TN 37235, US.
		Email addresses: muhammad.nadeem@vanderbilt.edu,  mirsaleh.bahavarnia@vanderbilt.edu, ahmad.taha@vanderbilt.edu. This work is supported by National Science Foundation under Grants 2152450 and 2151571.}
}

\begin{document}

	\newdimen\origiwspc%
	\newdimen\origiwstr%
	\origiwspc=\fontdimen2\font
	\origiwstr=\fontdimen3\font
	
	\fontdimen2\font=0.63ex
	
	\maketitle

	\markboth{IEEE Transactions on Power Systems, In Press, October 2023}{}

	\begin{abstract}
		Due to the rapid developments in synchronized measurement technologies, there exist enormous opportunities to attenuate disturbances in future power grids with high penetration of renewables and complex load demands. 
		To that end, this paper investigates the effectiveness of new robust feedback controllers for interconnected power systems with advanced power electronics-based models of photovoltaic (PV) power plants, composite load dynamics, and detailed higher-order synchronous generator models. Specifically, we design new, advanced control-theoretic  wide-area controllers to improve the transient stability of nonlinear differential-algebraic models. Thorough simulation studies are carried out to assess the performance of the proposed controllers. Several fundamental questions on the proposed controllers' computational complexity and disturbance attenuation performance are raised and addressed. Simulation results demonstrate that with the proposed controllers as a secondary control layer, the overall transient stability and system robustness against load and renewables disturbances/uncertainties can be significantly improved compared to the state-of-the-art.
	\end{abstract}
	\vspace{-0.1cm}
	\vspace{-0.17cm}
	
	\section{Introduction and motivation}\label{section:intro}
	\lettrine[lines=2]{T}{o} limit frequency nadir and the rate of change of frequency (RoCoF), and to bring a power system back to its steady-state/equilibrium conditions after a large disturbance, traditionally there have been three main control layers in the electrical grid: primary, secondary, and tertiary layers.  The primary control layer which commonly consists of a power system stabilizer (PSS), automatic voltage regulator (AVR), generator droop control, and generator inertial response, is usually responsible for regulating the frequency dynamics by providing damping to the system oscillation. The secondary control layer which consists of an automatic generation control (AGC), removes the steady-state error and tries to bring the system back to its nominal value, while the tertiary control layer is used for economic dispatch \cite{Anderson2003}. For decades, this control architecture has worked very well because of the flexibility of power generation from synchronous generators and the mostly predictive overall load demand curve. 
	
	However, with the massive penetration of complex load demands and uncertain solar/wind-based inverter-based resources (IBRs), power systems are facing significant challenges, and the overall transient stability and dynamic response of the system are deteriorating. This requires rethinking how the power system needs to be secured, protected, and operated \cite{NikosITPWRS2021,
		IEEtaskforce}. Unlike synchronous generators, IBRs are intermittent, they do not provide rotational inertia, and they are connected to the grid via power electronics-based technologies. Thus, this leads to increased frequency nadir, faster dynamical behavior, and system oscillations.

	To handle these uncertainties and transient stability-related issues in the future power grid, there exist enormous opportunities in real-time wide-area monitoring and control of power systems. This has been highly encouraged by the recent developments in synchronized measurement technologies in power systems and advanced robust state/output feedback controller designs that can handle the complex load demands and uncertain dynamics of renewables. These state/output feedback controllers can play a crucial role in the future interconnected power grids as they can send real-time control signals to the power plants based on the actual measurement received from the system  \cite{TerzijaIEEE2011, LiuITPWRS2021, NadeemITPWRS2022}.  Many studies have been carried out in the recent decade accordingly proposing centralized, localized, and distributed/decentralized feedback controllers.
	
	As an example of localized and distributed feedback controllers, researchers in \cite{RinaldiICSL2017} have designed a distributed $3^{rd}$-order sliding mode load frequency controller for a multi-machine power system model. Similarly, in \cite{Chuang}, a robust feedback controller for load frequency control based on  $\mathcal{H}_\infty$ stability has been proposed. In \cite{Zhao2014}, a decentralized frequency controller based on Lyapunov stability theory has been proposed for synchronous machines. In \cite{YuzhenITAC}, a decentralized controller based on the dissipative-Hamiltonian realization of the power system has been proposed for a multi-machine nonlinear differential algebraic equation (NDAE) model power system. 
	In \cite{SiljakITPWRS2002}, a linear matrix inequality (LMI) based controller derived via Lyapunov stability has been proposed for a highly simplified multi-machine model of a power system. This LMI-based application has later been extended in \cite{MarinoviciITPWRS2013} where $2^{nd}$-order generator model has been considered along with excitation and governor dynamics. 

	Centralized feedback control architectures have also been proposed extensively in the past decades, such as \cite{DorflerITPWRS2014} wherein an optimal wide-area controller (WAC) has been proposed to damp inter-area oscillations in a multi-machine power system model. Similarly, in \cite{ZhangITPWRS2013} a WAC for a linearized ordinary differential equation (ODE) model of a power system has been proposed to improve inter-area oscillations. In \cite{JainACC2015}, a model predictive control (MPC) based WAC has been proposed for a linearized and simplified power system model to improve inter-area oscillations. In \cite{ZolotasITCS2007}, a classical linear-quadratic Gaussian (LQG) based centralized feedback controller has been proposed for a multi-machine ODE-based power system model. In \cite{ChaudhuriITCS2012}, a classical linear-quadratic regulator (LQR) type WAC has been designed for a linearized ODE-based power system model to damp electromechanical oscillations. Recently, some learning-based centralized feedback controllers have also been presented, such as, in \cite{HadidiITSG2013}, a (reinforcement learning)-based WAC has been designed for a simplified multi-machine power system model. In \cite{YousefianITIA2016}, a temporal difference learning-based WAC has been proposed to improve the dynamic stability of the power system. Similarly, in \cite{YousefianPESGM2016}, a hybrid WAC based on Lyapunov stability and reinforcement learning has been designed for multi-machine power systems with wind energy resources.

	It is noteworthy that the majority of the studies in the aforementioned  feedback control of power systems mostly linearize the dynamics and consider a simplified representation of the electrical grid. This can cause serious stability issues as simplified controller designs may only work around the vicinity of the equilibrium point \cite{Nugroho_ITCST2023}. Also, algebraic constraints (power/current balance equations), dynamics of loads, and power electronics-based models of renewable energy resources are not taken into account  while designing a feedback controller. Furthermore, in the current literature, most of the feedback controllers are designed only for synchronous generators and very few feedback controllers have been proposed that can also control RERs in a real-time manner.
	
	Recently, some efforts have been carried out to address these limitations. In \cite{Nugroho_ITCST2023}, the authors propose a nonlinear feedback controller with algebraic constraints modeled in the controller architecture. However, a simplified $4^{th}$-order generator model is considered, uncertainties from load and renewables are not modeled in the controller design, load dynamics are ignored, and renewables are just considered as negative loads.
	Furthermore, in some recent studies feedback controllers for RERs have also been proposed such as \cite{StanojevITPWRS2022} in which MPC-based feedback controller has been designed to adjust the power set points of PV plants operating in grid-forming (GFM) mode so that they can support the grid after a large disturbance. However, the proposed feedback controller only considers the dynamics of the PV plant and sends control signals based on them, and the rest of the power system dynamics (dynamics of the network and synchronous machines) are completely neglected. Also, solving an MPC online in a real-time manner can highly be inefficient and require too much computational power.  A study that is close to the work presented here is \cite{SadamotoICSM2019} in which a two-layer (decentralized for wind/solar and centralized for synchronous machines) feedback control architecture has been proposed for multi-machine power system model with advanced (power electronics)-based models of wind farms, solar farms, and higher order synchronous generator model. However, power system algebraic constraints, load dynamics, and uncertainties from load and renewables are not considered in the controller architecture.
	
	We also want to mention here that, in this study GFM solar plants are used and secondary controller has been designed for them. The GFM strategy here is based on the conventional droop-controlled strategy \cite{ChandorkarITIA1993,lin2021stabilizing} and the solar plants are acting as a voltage sources and are regulating their terminal voltages similar to synchronous generators. The complete detailed explanation of the control strategy and dynamics of the solar power plant used in this study can be found in \cite{SoumyaITPWRS2022, WasynczukITPE1996}. On the other hand, grid-following (GFL) solar power plants are designed to simply follow grid voltage and frequency generated by synchronous generators and thus they also commonly require modeling of PLL dynamics in their design \cite{KenyonNREL}. The overall difference between the control strategy of the GFM PV plant model used in this study and the GFL model can be visualized from their block diagrams given in Fig. 1 and 2 of \cite{KenyonNREL}. Notice that GFM designs are more preferred because of their black start capabilities and ability to independently regulate their voltage and frequency \cite{lin2021stabilizing}.
	
	\noindent \textbf{Paper Contributions. } In the light of the above discussion and aforementioned limitations, in this work, we propose a robust wide-area controller for a highly interconnected power system model with a power electronics-based model of PV plants, composite load dynamics, and comprehensive $9^{th}$-order synchronous generator dynamics. The proposed wide-area controller sends additional control signals not only to synchronous generators but also to PV power plants to adjust their power outputs after large disturbances so that the power system remains stable and the overall transient stability can be improved. The key paper contributions are as follows:
	
	\begin{itemize}
		\item This is the first work to propose a feedback controller for an advanced interconnected power system model. The system considered here comprises (a) advanced (power electronics)-based models of PV plants, (b) comprehensive $9^{th}$-order synchronous machine dynamics, (c) motor loads, (d) constant power loads, and (e) constant impedance loads. In addition to synchronous generators, the proposed feedback controller here also sends control signals to PV power plants and actively adjusts their power output during a disturbance so that the overall transient stability of the system can be improved.
		\item The proposed controller design explicitly models the algebraic constraints of the power systems and also considers nonlinearity in the controller architecture by modeling it as an $\mathcal{L}_2$-norm bounded uncertainty, which is more realistic as compared to completely neglecting it through linearization. 
		
		\item To handle uncertainty from load and renewables, we utilize robust $\mathcal{H}_\infty$ notion in the controller design. The main advantage of  $\mathcal{H}_\infty$ based controller design is that it does not require any statistical knowledge of the disturbance and can minimize the impact of any sort of bounded disturbance on the system dynamics \cite{NadeemITPWRS2022}. 
		
		\item The performance of the proposed feedback controller has been analyzed on IEEE $9$-bus and $39$-bus test systems by running extensive simulation studies. Specifically, (a) we assess the performance of the proposed robust WAC under different severity of disturbances from load and renewables. The advantages of the proposed feedback controller are also presented by comparing the dynamic response of the power network with conventional control (primary controllers of power systems) and with robust WAC acting on top of them,  (b) we also reduced the initial complete NDAE power system model to an NODE system and the simplified versions of the proposed WAC have also been designed based on the NODE system model. Then, the overall performance and computational efficiency among them have been thoroughly discussed.  	
	\end{itemize}
	\noindent {\textbf{Notations:}}
	The sets are represented in calligraphic such as $\mathcal{G}, \mathcal{U}$, etc. All the vectors and matrices are bold-faced.  The notation $\m I$ denotes an identity matrix while $\m O$ represents a zero matrix of appropriate dimensions. The notation $\mathbb{R}^{x}$ denotes a row vector with $x$ elements. The notation $\mathbb{R}^{x\times y}$ denotes a real matrix of size $x$-by-$y$. Similarly $\mathbb{S}^{x\times y}_{++}$ denotes a positive definite matrix of size $x$-by-$y$. The symbol $*$ represents symmetric entries in a symmetric matrix. We denote the Frobenius norm of a matrix $M$ by $\|M\|_F$.
	\vspace{-0.1cm}
	\section{PV and load-integrated system model}\label{sec: System model}
	We consider a power system model with $G$ synchronous generators, $R$ solar power plants, and $L_z$, $L_p$, $L_k$ number of constant impedance, constant power, and motor loads, respectively.  The overall power system is represented as a graph ($\mathcal{N}, \mathcal{E}$),  where $\mathcal{N} = \left\lbrace 1,\dots,N\right\rbrace$ denotes the set of buses and $\mathcal{E}$ represents the set of transmission lines. Notice that $\mathcal{N} = \mathcal{G}\cup \mathcal{R}\cup\mathcal{U}\cup\mathcal{L}$, where $\mathcal{G} = \left\lbrace 1,\dots,G\right\rbrace$ denotes the set of buses connected to synchronous generators, $\mathcal{R} = \left\lbrace 1,\dots,R\right\rbrace$ represents set of buses with solar power plants, $\mathcal{U}$ represents set of non-unit buses, and $\mathcal{L}$ collects set of buses connected to $L_z$, $L_p$, and $L_k$.  
	
	We model the power system as a set of nonlinear differential equations (detailing the dynamic models of synchronous generators, grid-forming PV plants, and dynamic loads) and algebraic equations (describing power/current balance equations) as follows:
	\begin{subequations}~\label{equ:PSModel}
		\begin{align}
			\vspace{-0.5cm}
			\textit{Differential equations:} \;\;\;\;\;	\dot{\m x}(t) &= \m f(\m x_d,\m x_a, \m u, \m w) ~\label{equ:PSModel-a} \\
			\textit{Algebraic equations:} \;\;\;\;\;	\m 0 &= \m h(\m x_d,\m x_a, \m w) ~\label{equ:PSModel-b}
		\end{align}
	\end{subequations}
	where $\m x_d \in \mbb{R}^{n_d}$ represents dynamic variables and it lumps the states of generators, PV plants, and loads, $\m x_a \in \mbb{R}^{n_a}$ represents algebraic variables of the power network,  $\m u \in \mbb{R}^{n_u}$ models the control inputs and steers the system to its equilibrium after a disturbance, and $\m w \in \mbb{R}^{n_w}$ denotes exogenous disturbances such as uncertainties in load demand and solar irradiance.
	
	
	In  \eqref{equ:PSModel}, vector $\m x_a$ is modeled as:
	\begin{align}\label{eq:x_a}
		\m x_a :=  \m x_a(t) = \bmat{\m I_{\mr{Re}}^\top&\m I_{\mr{Im}}^\top&\m V_{\mr{Re}}^\top&\m V_{\mr{Im}}^\top}^\top \in \mbb{R}^{n_a}
	\end{align}
	where $\m I_{\mr{Re}}\hspace{-0.1cm}= \hspace{-0.1cm}\{I_{\mr{Re}_i}\}_{i\in \mc{N}}, \m I_{\mr{Im}}\hspace{-0.1cm}= \hspace{-0.1cm}\{I_{\mr{Im}_i}\}_{i\in \mc{N}}, \m V_{\mr{Re}}\hspace{-0.1cm}= \hspace{-0.051cm}\{V_{\mr{Re}_i}\}_{i\in \mc{N}}, \m V_{\mr{Im}}\hspace{-0.05cm}=\hspace{-0.05cm} \hspace{-0.05cm}\{V_{\mr{Im}_i}\}_{i\in \mc{N}}$ represent the real and imaginary parts of current and voltages, respectively. The vector $\m{u}$ models the control inputs of synchronous generators and solar PV plants and is represented as
	\vspace{-0.1cm}
	\begin{align}
		\m u :=\m u(t) = \bmat{\m u_G^\top & \m u_R^\top}^\top\in \mbb{R}^{n_u}
	\end{align}
	where $\m u_G = \bmat{\m V_g^{*\top} & \m P_v^{*\top}}^\top \in \mbb{R}^{2G}$ with  $\m V_g^*$ and $\m P_v^*$ denoting reference set-points for voltages $\mr{(pu)}$ and  turbine valve positions $\mr{(pu)}$  of the synchronous generator, respectively. Similarly, $ \m u_R = \bmat{\m V_s^{*\top} & \m P_s^{*\top}}^\top \in \mbb{R}^{2R}$, where $\m P_s^*$ and $\m V_s^*$ are the power $\mr{(pu)}$ and voltage $\mr{(pu)}$ reference set points for solar PV plants. Also, we define $\m w$ in Eq. \eqref{equ:PSModel} as $\m w = \bmat{\m I_{r}^\top&\m P_{d}^\top}^\top \in \mbb{R}^{n_w}$ where $ \m I_r$ is the solar irradiance $(W/m^2)$ on the PV plants and $ \m P_d$ is the system real power load demand $\mr{(pu)}$.

	Moreover, in Eq. \eqref{equ:PSModel}, we represent $\m x_d$ as
	\begin{align}\label{eq:x_d}
		\m x_d :=\m x_d(t)= \bmat{\m x_G^\top&\m x_R^\top&\m x_m^\top}^\top \in \mbb{R}^{n_d}
	\end{align} 
	where $\m x_G$ are the dynamic states of the conventional power plant (states of synchronous generator, excitation system, governor, and turbine dynamics), $\m x_R$ represents the dynamic states of the solar power plant, and  $\m x_m$ denotes the states of motor loads. We model the conventional power plant via a comprehensive $9^{th}$-order model, and thus vector $\m x_G$ can be expressed as follows \cite{sauer2017power,SoumyaITPWRS2022}:
	\begin{align*}
		\m x_G\hspace{-0.01cm} = \hspace{-0.01cm}\bmat{\m \delta_{\mr{g}}^\top\;\;\;\m \omega_{\mr{g}}^\top\;\;\m E_{\mr q}^\top\;\;\; \m E_{\mr d}^\top\,\,\m T_\mr{M}^\top\,\,\m P_{v}^\top\,\,\m E_{\mr{fd}}^\top\;\;\m r_{f}^\top\;\; \m v_{a}^\top}^\top \hspace{-0.2cm}\in\hspace{-0.01cm} \mbb{R}^{9G} \label{eq:stateSyncGen}
	\end{align*}
	where $\m \delta_{\mr g}$ denotes generator rotor angle $\mr{(pu)}$, $\m \omega_{\mr g}$ is the generator speed $\mr{(pu)}$, $\m E_{\mr q}$, $\m E_{\mr d}$, represent transient voltages along dq-axis $\mr{(pu)}$, $\m T_{\mr M}$ denotes turbines prime mover torque $\mr{(pu)}$, $\m P_{v}$ is the turbine valve position $\mr{(pu)}$,   $\m E_{\mr{fd}}$ is the generator field voltage $\mr{(pu)}$, $\m r_{f}$ denotes stabilizer output $\mr{(pu)}$, and $\m v_{a}$ represents amplifier voltage $\mr{(pu)}$. Readers are referred to Appendix \ref{appndix:ninth Gen_dynamics} for the detailed description of synchronous generator dynamics. 
	
	We leverage the $12^{th}$-order grid-forming PV plant model $i \in$ $\mathcal{R}$ in \cite{SoumyaITPWRS2022, WasynczukITPE1996}. 
	The overall model describes, DC side dynamics (DC link and PV array dynamics), AC side dynamics (DC/AC converter and LCL filter dynamics), and voltage/current regulators models, thus the state vector $\m x_R$ for solar plants  can be written as: 
	\begin{align}
		\small \hspace{-0.3cm}	\m x_R\hspace{-0.01cm} =\hspace{-0.01cm} \bmat{\m E_{\mr{dc}}^\top\;\;\;\m i_{\mr{dqf}}^\top\;\;\;\m v_{\mr{dqc}}^\top \;\;\;\m \delta_{\mathrm{c}}^\top\;\;\;\m P_{e}^\top\;\;\;\m Q_{e}^\top\;\;\;\m z_{\mr{dqo}}^\top\;\;\;\m z_{\mr{dqf}}^\top}^\top\hspace{-0.15cm} \in \hspace{-0.05cm}\mbb{R}^{12R}
	\end{align}
	where  $\m{E}_{\mathrm{dc}}$ is the energy stored in the DC side capacitor, $\m{i}_{\mr{dqf}}=[\m i_{d_f}^\top\,\,\m i_{q_f}^\top]$ represent the currents $\mr{(pu)}$ at the terminals of the inverter along dq-axis, $\m{v}_{\mr{dqc}} \hspace{-0.05cm}=\hspace{-0.02cm} [\m v_{d_c}^\top\,\,\m v_{q_c}^\top]$ are the voltages $\mr{(pu)}$ across the AC capacitor along dq-axis,  $\m{\delta}_{{c}}$ represents the  inverter angle $\mr{(pu)}$ of solar power plants, $\m{{P}}_{e}$, $\m{{Q}}_{e}$ are the total real and reactive power injected by solar plants to the grid, and $\m{z}_{\mr{dqo}}\hspace{-0.1cm}=\hspace{-0.04cm}[\m z_{d_0}^\top\,\,\m z_{q_0}^\top]$, $\m{z}_{\mr{dqf}}\hspace{-0.01cm}=\hspace{-0.04cm}[\m z_{d_f}^\top\,\,\m z_{q_f}^\top]$ are the dynamic states of the voltage and current regulator of PV plants. To obtain further information and a detailed description of the solar power plant model used in this study, readers are referred to Appendix \ref{appndix:Grid forming PV dynamics}.
	
	The dynamical model \eqref{eq:x_d} for the motor loads is defined as \cite{krause2013}:
	\vspace{-0.12cm}
	\begin{equation}
		\dot{\omega}_{\mr{m}_i} = \frac{1}{2H_{\mr m_i}}(T_{{e_i}} - T_{\mr m_i}) \label{eq:omegam}
	\end{equation}
	where $\omega_{\mr m_i}$ is the motor speed, $H_{\mr m_i}$ is the motor inertia constant, and $T_{{m_i}}$, $T_{{e_i}}$, represent mechanical and electromagnetic torques of the motor, respectively \cite{krause2013}. 
	Furthermore, constant power and constant impedance loads satisfy the following relationships \cite{NERC}: 
	\begin{subequations}
		\begin{align}
			\mr{conj}(I_{p_i})V_{p_i}+(P_{p_i}+Q_{p_i}) &= 0\\
			\begin{split}\label{eq:load_dyn_z}
				V_{z_i} + I_{z_i}Z_i &= 0
			\end{split}
		\end{align}
	\end{subequations}
	where $	\mr{conj}$ represents complex conjugate operator, $I_{p_i}$, $V_{p_i}$ are the current and voltage phasor of bus connected to constant power loads, and $P_{p_i}$, $Q_{p_i}$ are the real and reactive power of constant power loads, respectively. Similarly, in \eqref{eq:load_dyn_z}, $V_{z_i}$, $I_{z_i}$ are the current and voltage phasor of the bus connected to a constant impedance load $Z_i$. This completes the modeling of system dynamics \eqref{equ:PSModel-a}.
	
	Now, the model for the algebraic constraints Eq. \eqref{equ:PSModel-b} is expressed as \cite{sauer2017power}: 
	\begin{gather}
		\underbrace{\begin{bmatrix}
				\m{{I}}_{G} \\ \m{{I}}_{R} \\ \m{{I}}_{L}
		\end{bmatrix}}_{\m I(t)}
		-
		\underbrace{\begin{bmatrix}
				\m{Y}_{GG} & \m{Y}_{GR} & \m{Y}_{GL} \\
				\m{Y}_{RG} & \m{Y}_{RR} & \m{Y}_{RL} \\
				\m{Y}_{LG} & \m{Y}_{LR} & \m{Y}_{LL} 
		\end{bmatrix}}_{\m Y}
		\underbrace{\begin{bmatrix}
				\m{{V}}_{G} \\ \m{{V}}_{R} \\ \m{{V}}_{L} \\
		\end{bmatrix}}_{\m V(t)} =  \m{0} \label{eq:transalgebraic}
	\end{gather}
	where $\m V$ denotes bus voltages, $\m I$ represents net injected current, and $\m{Y}$ is the network admittance matrix. In \eqref{eq:transalgebraic}, $\m{{V}}_{R} \hspace{-0.03cm}= \hspace{-0.03cm}\{V_{Re_i}\}_{i\in \mc{R}}\hspace{-0.05cm}+\hspace{-0.05cm} j\{V_{Im_i}\}_{i\in \mc{R}}\hspace{-0.05cm}$ denotes voltage phasors at the terminal of buses connected with PV power plant and $\m{{I}}_{R}\hspace{-0.01cm}= \hspace{-0.03cm}\{I_{Re_i}\}_{i\in \mc{R}}\hspace{-0.05cm}+\hspace{-0.05cm}j \{I_{Im_i}\}_{i\in \mc{R}}\hspace{-0.05cm}$ represents current phasors injected by the PV power plant. Similarly  
	$\m{{I}}_{G}$, $\m{{I}}_{L}$, and $\m{{V}}_{G}$, $\m{{V}}_{L}$  denote current and voltage phasors of all loads and synchronous generators, respectively. 
	
	That being said, by considering equations (\ref{eq:x_a})--(\ref{eq:transalgebraic}) and including the associated dynamic models given in Appendices \ref{appndix:ninth Gen_dynamics} and \ref{appndix:Grid forming PV dynamics}, we can express the overall interconnected model of the power system in a compact state-space format as: 
	\begin{align}\label{eq:final_NDAE}
		\hspace{-0.21cm}\boxed{	\mr{\textbf{NDAE:}}\;\;\,\m E\dot{{\m x}} = {\m A}{\m x}\hspace{-0.03cm} +\hspace{-0.03cm}  {\m f}\left({\m x},{\m u},{\m w} \right) + {\m B} {{\m u} } + {\m B}_w \m w}
	\end{align}
	where $\m E$ is a binary singular matrix and it encodes algebraic equations with rows of zeros and $\m x(t) = \bmat{\m x_d^\top & \m x_a^\top}^\top \in\mbb{R}^{n}$ represents the overall state vector.  The constant state-space matrices $\m A, \m B$, and $\m B_w$ in \eqref{eq:final_NDAE} are computed via capturing the linear components of NDAE model \eqref{equ:PSModel} while the function ${\m f}\left({\m x},{\m u},{\m w} \right)$ represents the encompassed nonlinearities. 
	
	Notice that \eqref{eq:final_NDAE} models the complete NDAE representation of the power system without any assumptions. Ideally, the feedback controller should be designed for this complete NDAE representation of the power system without any simplifications so that the controller has knowledge about all the uncertainties and nonlinearities to provide robust performance. However, in literature commonly the algebraic variables are usually eliminated by converting the system to a nonlinear ODE (NODE) model. This is mainly because of the better understanding and rich control theoretic literature about the NODE system as compared to NDAE models. Furthermore, it has been shown that NODE-based controller designs are also effective in controlling 
	the initial complete NDAE system \cite{TahaITCNS2019, BazrafshanITSG2019}. However, further study is required to compare the performance of both types of controller architectures.  In this work, we design a controller based on both the complete NDAE representation of power system \eqref{eq:final_NDAE} and also its NODE counterpart given below. 
	
	Notice that in both types of controller design, the final implementation is done on the complete  power system model without any simplifications or assumptions. In the NODE-based controller design, the controller takes information only from NODE system matrices while in the NDAE-based controller, the controller utilizes information from NDAE state-space system matrices. 
	
	With that in mind, considering 
	\begin{align*}
		\m x &= \bmat{\m x_d^\top & \m x_a^\top}^\top,~
		\m A = \bmat{\m A_{dd}& \m A_{da}\\ \m A_{ad} & \m A_{aa}}\\ \m B &= \bmat{\m B_{d}^\top&\m B_{a}^\top}^\top,~\m B_w = \bmat{\m B_{wd}^\top&\m B_{wa}^\top}^\top \end{align*}
	and assuming $\m A_{aa}$ to be invertible (which is common in the area of power systems---see\cite{TahaITCNS2019,AranyaICSM2019}), then the algebraic variables $\m x_a$ can be eliminated from \eqref{eq:final_NDAE} and the equivalent NODE representation of model \eqref{eq:final_NDAE} can be written  as 
	\begin{align}\label{eq:final_NODE}
		\hspace{-0.2cm}\boxed{	\mr{\textbf{NODE:}}\;\;\, \dot{\m x}_d \hspace{-0.02cm}=\hspace{-0.02cm} \tilde{\m A}{\m x_d} \hspace{-0.02cm}+\hspace{-0.02cm} {\m f_d}\left({\m x_d},{\m u},{\m w} \right) \hspace{-0.02cm}+\hspace{-0.02cm} \tilde{\m B} {{\m u} } \hspace{-0.02cm}+\hspace{-0.02cm}\Bar{\m B}_w \m w}
	\end{align}
	where ${\m f_d}$ is the corresponding nonlinear mapping and the rest of the constant matrices $\tilde{\m A}$, $\tilde{\m B}$, and $\Bar{\m B}_w$ are given as follows:
	\begin{align*}
		\tilde{\m A} &= \m A_{dd} - \m A_{da}\m A_{aa}^{-1}\m A_{ad},\;\;\;   \tilde{\m B} = \m B_{d} - \m A_{da}\m A_{aa}^{-1}\m B_{a}\\
		\Bar{\m B}_w &= \m B_{wd} - \m A_{da}\m A_{aa}^{-1}\m B_{wa}.
	\end{align*}
	In the following section, we discuss $\mathcal{H}_\infty$ stability criterion and present the architecture of the proposed wide-area controller based on both NDAE \eqref{eq:final_NDAE} and NODE \eqref{eq:final_NODE} systems.
	\vspace{-0.2cm}
	\section{Robust Feedback Controller Designs}\label{sec:controller design}
	In this section, we present a host of novel wide-area controllers for both NDAE \eqref{eq:final_NDAE} and NODE \eqref{eq:final_NODE} models with varying properties. In particular, our proposed WACs are of three different types: (1)  $\mathcal{H}_\infty$ NDAE-based, (2)  $\mathcal{H}_\infty$ NODE-based, and (3) $\mathcal{H}_2$ NODE-based feedback controller designs. 
	\vspace{-0.4cm}
	\subsection{WAC based on NDAE System Model}
	Here we design an LMI-based wide-area robust state feedback controller for an interconnected model of a power system having $9^{th}$-order generator dynamics, grid-forming PV plant model, and composite load dynamics as detailed in Section \ref{sec: System model}. The overall objective of the controller design is to improve the transient stability of the system by providing damping to the frequency oscillation after a large fault/disturbance. The proposed controller act as a secondary control layer and provides additional control signals to the primary control layer of the interconnected power network which comprises of PSSs, AVRs (for the synchronous generators), and PI controllers (voltage and current regulators) of grid-forming PV power plants. To that end, the power system model \eqref{eq:final_NDAE} with the proposed controller (the closed-loop system) can be written as follows:
	\begin{align}\label{eq:final_NDAE_cntrl}
		\m E\dot{{\m x}} &= {\m A}{\m x} +  {\m f}\left({\m x},{\m u_{cl}},{\m w} \right) + {\m B} {{\m u_{cl}} } + {\m B}_w \m w
	\end{align}
	in which the closed-loop state feedback control input $\m u_{cl}$ for $kT \leq t< (k+1)T$ is given as:
	$$\boxed{{\m u}_{cl}:= \m u_{cl}(t) = \m u_{{ref}}^k(t) + \m K\left({\m x(t)} - \m x^k(t)\right)}$$
	where $\m u_{{ref}}^k$ is the reference set point of the control input $\m u$ which is commonly determined for every $k^{th}$-dispatch time-period by running power flow (PF) or optimal power flow (OPF), $\m x^k$ is the steady state value of the state vector before the occurrence of disturbances/fault, and $\mK \in\mbb{R}^{n_u\times n} $ is the controller gain matrix. The main objective of this paper is to design a controller gain matrix $\mK$ such that after a fault/disturbance occurrence, the controller minimizes the impact of fault/disturbance on the state dynamics via the state feedback information and consequently improves the transient stability of the system. 
	
	To proceed with the computation of $\mK$, let us assume there is a large perturbation caused by  fault/disturbance in load/renewables and the new steady-state value of vector $\m w$ is $\m w^e$. This fault/disturbance will eventually push the system states to a new equilibrium, let us denote that by  $\m x^e$. Then, the system dynamics \eqref{eq:final_NDAE_cntrl} with the proposed controller at this new equilibrium point can be rewritten as: 
	\begin{align*}
		\m 0 &= \m A{\m x^e}\hspace{-0.05cm} + \hspace{-0.05cm}\m f\left(\m x^e,\m u_{cl},\m w^e \right)\hspace{-0.05cm} +\hspace{-0.05cm} \m B(\m u_{{ref}}^k\hspace{-0.05cm} + \hspace{-0.05cm}\m K ({\m x^e}\hspace{-0.05cm} - \hspace{-0.05cm}\m x^k)) \hspace{-0.05cm}+\hspace{-0.05cm} \m B_w \m w^e.
	\end{align*}
	
	With that in mind, to analyze the system behavior after the fault, let us define $\Delta \m x = \m x-\m x^e$ and $\Delta \m w = \m w-\m w^e$ as the \textit{deviations} around the new equilibrium $({\m x}^e,{\m w}^e)$, respectively. Then the perturbed closed-loop system dynamics can be expressed as follows: 
	\begin{align}\label{eq:final_NDAE_peturbed}
		\m E\Delta\dot{\m x}\hspace{-0.05cm} &=\hspace{-0.05cm} (\m A\hspace{-0.05cm}+\hspace{-0.05cm}\m{BK})\Delta\m x+\Delta\m f(\Delta \m x, \m u_{cl}, \Delta \m w)\hspace{-0.05cm} + \hspace{-0.05cm}\m B_w \Delta\m w
	\end{align}
	where $\Delta\m f(\Delta \m x, \m u_{cl}, \Delta \m w) = \m f(\m x,\m u_{cl},\m w)-\m f(\m x^e,\m u_{cl},\m w^e)$. Notice that $\Delta\m w$ represents the deviation of load demand and solar irradiance from their respective steady-state values $\m w^e$.
	The objective of the controller gain $\mK$ is to drive all the solution trajectories of the NDAE model \eqref{eq:final_NDAE_peturbed} to zero and attenuate the impact of uncertainty $\Delta\m w$ on the power system dynamics which is equivalent to saying that the controller ensures that the power system model \eqref{eq:final_NDAE_cntrl}  asymptotically converges to the new equilibrium point ($\m x^e$, $\m w^e$) after a fault/disturbance occurrence. We now discuss $\mathcal{H}_\infty$ stability criterion and present the theory of the proposed controller which is posed as a convex semi-definite program (SDP). 
	
	$\mathcal{H}_\infty$ control is a well-established and powerful mathematical tool in modern control theory literature. It provides stability and guaranteed performance (for a particular $\mathcal{H}_\infty$ criterion) of the dynamical system under large uncertainties/disturbances. The basic idea in $\mathcal{H}_\infty$ control is that first a performance criterion for the control law is considered. Then, via a state feedback architecture, a controller gain is determined subject to the attenuation of the impact of disturbances on the designed performance criterion \cite{masubuchi1997h}. 
	
	To that end, let us assume $\m z_1 := \m z_1(t) = \mC\m x(t) + \mD\m u_{cl}(t) + \mD_w\m w(t) \in\mbb{R}^{n}$ be the performance index of the control law $\m u_{cl}$, where $\mC \in\mbb{R}^{n\times n}$, $\mD \in\mbb{R}^{n\times n_u}$, and $\mD_w \in\mbb{R}^{n\times n_w}$ are constant penalizing matrices. Similar to the matrices $\mQ$ and $\mR$ in the vintage LQR control, matrices $\mC$, $\mD$, and $\mD_w$ can be determined based on the grid operator preferences, meaning how much and which state or control input should be penalized while designing the controller gain $\mK$. Now similar to as done previously the perturbed performance index around equilibrium $\m x^e$ can be written as $\Delta \m z_1= \m z_1-\m z_1^e = (\m C+\m{DK})\Delta\m x + \mD_w\Delta\m w$. 
	
	With that in mind, for the sake of notation simplicity, from now on, with a little abuse of notation, let  $\Delta\m x = \m x$, $\Delta\m f(\Delta \m x, \m u_{cl}, \Delta \m w)= \m f(\m x,\m u_{cl},\m w)$, $\Delta\m z_1 = \m z_1$, and $\Delta\m w = \m w$. Then, the perturbed closed-loop dynamics \eqref{eq:final_NDAE_peturbed} with performance index $\m z_1$ can be rewritten as:
	\begin{subequations}\label{eq:final_NDAE_peturbed_final}
		\begin{align}
			\m E\dot{\m x} &= (\m A+\m{BK})\m x+\m f(\m x,\m u_{cl},\m w)+ \m B_w \m w\\
			\m z_1 &= (\m C+\m{DK})\m x + \mD_w\m w.
		\end{align}
	\end{subequations}
	To that end, the $\mathcal{H}_\infty$ stability criterion can be expressed as follows:
	\begin{mydef}\label{def:H_inf}
		The performance $\m z_1$ of the control action $\m u_{cl}$ is robust in the sense of $\mathcal{H}_\infty$ with performance level $\mu$ if, $(a)$ the perturbed system dynamics \eqref{eq:final_NDAE_peturbed_final} are asymptotically stable when $\m w = \m0$ for all $t>0$ and $(b)$ $\norm{\m z_1}^2_{\mathcal{L}_2} <  \mu^2\norm{{\m w }}^2_{\mathcal{L}_2}$ for zero initial perturbation (e.g., $\m x(0)=\m0$) and for any $\mathcal{L}_2$-norm bounded uncertainty $\m w$.
	\end{mydef}
	
	Definition \ref{def:H_inf} can be interpreted as follows: when disturbance $\m w = \m0$ for all $t>0$ then the power system rests in a steady-state status and is thus stable. However, when there is some unknown $\mathcal{L}_2$-norm bounded uncertainty $\m w$, then, $\mathcal{H}_\infty$ stability guarantees that the magnitude of the performance output $\m z_1$ of the closed-loop dynamics always evolves in a way such that it is less than $\mu$ times the magnitude of uncertainty $\m w$, where $\mu$ here is an optimization variable and is commonly called as the performance level of $\m z_1$. In $\mathcal{H}_\infty$-based controllers, one tries to minimize $\mu$ to get robust performance from the controller under various sources of disturbances.
	
	We now present a systematic approach, based on Lyapunov stability theory, to synthesize a controller gain $\mK$ that guarantees $\mathcal{H}_\infty$ stability for the closed-loop dynamics \eqref{eq:final_NDAE_peturbed_final} under disturbance $\m w$. Before that, we consider the following assumption throughout the paper:
	\textcolor{black}{\begin{asmp}\label{asmp:regular}
			The pair $(\mE,\mA)$ is regular and the triplet $(\mE,\mA, \mB)$ is finite dynamics stabilizable and impulse controllable.
		\end{asmp}The above assumption is standard in control theoretic literature \cite{TakabaCDC,cobb1984controllability}, and various power system models have been showcased to be indeed regular and stabilizable---see \cite{AranyaICSM2019,Nugroho_ITCST2023}. To that end, we now present the following main result.} 
	\begin{theorem}\label{theorm:H_inf}
		{Suppose that Assumption \ref{asmp:regular} holds.} Then, the perturbed closed-loop dynamics \eqref{eq:final_NDAE_peturbed_final} with performance index $\m z_1$ is $\mathcal{H}_\infty$ stable if there exists matrices $\m X \in \mbb{S}_{++}^{n \times n}$,  $\mW\in \mbb{R}^{n_a\times n}$, $\mH\in \mbb{R}^{n_u\times n}$ and a scalar $\lambda\in\mbb{R}_{++} $ such that the following convex semi-definite optimization program is feasible    
		\begin{align*}
			\mathbf{\left( OP_1\right) }\;\;\;\;\;\minimize_{{\lambda}, \m H, \m X,\m W} &\;\;\; \lambda\\ \subjectto  & \;\;\;\mr{LMI}\; \eqref{eq:LMI_Hinf},\;\m X \succ \mO,\;\lambda>0
		\end{align*} 
		where $\mr{LMI}$ \eqref{eq:LMI_Hinf} is as follows:
		\begin{align}\label{eq:LMI_Hinf}
			\hspace{-0.1cm}\bmat{ \m\Psi & * & * \\ \tilde{\mB}_w^\top & -\lambda\m I &* \\
				\mC^\top(\mX\mE^\top+\mE^\perp\mW)+\mD\mH&\hat{\mD}_w&-\m I} \prec \mO
		\end{align}
		and $\m\Psi$ is given as:
		\begin{equation*}
			\scriptsize	\m\Psi\hspace{-0.01cm} =\hspace{-0.01cm} (\mX\mE^\top\hspace{-0.03cm}+\hspace{-0.03cm}\mE^\perp\mW)^\top\mA^\top \hspace{-0.03cm}+\hspace{-0.03cm} \mA(\mX\mE^\top\hspace{-0.03cm}+\hspace{-0.03cm}\mE^\perp\mW)\hspace{-0.03cm} +\hspace{-0.03cm} \mH^\top\mB^\top \hspace{-0.03cm}+\hspace{-0.03cm} \mB\mH
		\end{equation*}
		where $\mE^{\perp}\in\mbb{R}^{n\times n_a}$ is the orthogonal complement of $\mE$. Upon solving $\mathbf{OP_1}$ the controller gain can be computed as $\m K = \mH(\m X\m E^\top+\m E^{\perp}\mW)^{-1}$.
	\end{theorem}
	Readers are referred to Appendix \ref{appndix:Proof therm1} for the complete proof of Theorem \ref{theorm:H_inf}. We name the controller that is based on solving $\mathbf{OP_1}$ as $\mathcal{H}_\infty$-DAE controller. In Theorem \ref{theorm:H_inf}, we pose the controller design as a convex semi-definite optimization problem and thus, can easily be solved via commercial optimization solvers such as MOSEK \cite{Andersen2000}. 
	The calculated controller gain $\mK$ guarantees that the deviation in the magnitude of performance index $\m z_1$ is robust in the sense of $\mathcal{H}_\infty$ as discussed in Definition \ref{def:H_inf}, while also satisfying the budget constraint on the control inputs (which is enforced via matrices $\mC$ and $\mD$) or in other words, it ensures that $\norm{\m z_1}^2_{\mathcal{L}_2} $ always lies within a tube having an origin at zero and a radius of $\mu^2\norm{{\tilde{\m w} }}^2_{\mathcal{L}_2}$. Theorem \ref{theorm:H_inf} also makes sure that after a large disturbance, the closed-loop interconnected NDAE model of power system \eqref{eq:final_NDAE_cntrl} is asymptotically stable as $t\rightarrow\infty$ and the system dynamics converges to a new steady-state value $\m x_e$.  
	
	Additionally and in comparison to the method in \cite{AranyaICSM2019}, which proposes a feedback controller for highly interconnected power systems model, Theorem \ref{theorm:H_inf} includes the explicit modeling of algebraic equations and uncertainty from vector $\tilde{\m w}$ in the controller design, composite load dynamics, 
	and an SDP optimization routine (given above in $\mathbf{OP_1}$) that seeks to obtain
	an optimal solution to the design of controller gain $\mK$. In comparison, the method in \cite{AranyaICSM2019} only seeks a feasible solution.

	In the following section, we propose a variety of other state feedback controllers based on the NODE system model representation \eqref{eq:final_NODE}. 
	
	\subsection{WAC based on NODE System Model}
	Since a rich control theoretic literature exists for ODE systems, a variety of robust state feedback controllers can be designed based on the state-space model \eqref{eq:final_NODE}. To that end, we first design the perturbed dynamics for model \eqref{eq:final_NODE}. Similar to as done in the previous section, assuming that the perturbation in nonlinear function in the closed-loop dynamics \eqref{eq:final_NDAE_peturbed_final} is $\mathcal{L}_2$-norm bounded and can be written as $\Delta\m f(\m x,\m u_{cl},\m w) = \mB_f\m w_f$, where $\m B_f = \mB_w$ and by defining $\tilde{\mB}_w =\bmat{\Bar{\mB}_w & \m B_f}$, then 
	the perturbed closed-loop dynamics of the NODE model \eqref{eq:final_NODE} can be expressed as follows:
	\begin{align}\label{eq:NODE_peturbed}
		\dot{\m x}_d &= (\tilde{\m A}+\tilde{\m B}{\m K}_d)\m x_d+ \tilde{\m B}_w \tilde{\m w}.
	\end{align}
	Now, in a similar fashion by considering $\mC = \bmat{\mC_d & \mC_a}$, we can eliminate $\m x_a$ from the performance index $\m z_1$ and thus, the equivalent ODE representation of $\m z_1$, namely $\m z_2\in \mbb{R}^{n_d}$, can also be written as:
	\begin{align*}
		\m z_2 &= (\tilde{\m C}+\tilde{\m{D}}\m{K}_d)\m x_d + \Bar{\m D}_w{\m w}
	\end{align*}
	where
	\begin{align*}
		\tilde{\m C} &= \m C_{d} - \m C_{a}\m A_{aa}^{-1}\m A_{ad}, \;\; \tilde{\m D} = \m B_{d} - \m C_{a}\m A_{aa}^{-1}\m B_{a}\\
		\Bar{\m D}_w &= \m D_{w} - \m C_{a}\m A_{aa}^{-1}\m B_{wa}.
	\end{align*}
	To that end, the overall perturbed NODE system dynamics along with its performance index can be expressed as follows:
	\begin{subequations}\label{eq:final_NODE_peturbed}
		\begin{align}
			\dot{\m x}_d &= (\tilde{\m A}+\tilde{\m B}{\m K}_d)\m x_d+ \tilde{\m B}_w \tilde{\m w}\\
			\m z_2 &= (\tilde{\m C}+\tilde{\m{D}}\m{K}_d)\m x_d + \tilde{\m D}_w\tilde{\m w}
		\end{align}
	\end{subequations}
	where $\tilde{\m D}_w = \bmat{\Bar{\mD}_w& \mD_f}$ with $\mD_f = \mB_f$.
	
	Now, based on the perturbed dynamics \eqref{eq:final_NODE_peturbed} in this work, we design the following types of state feedback  controllers: 
	
	\subsubsection{{H-infinity-based ODE state-feedback controller design}}
	
	Here, we design an $\mathcal{H}_\infty$-based ODE  controller to determine the feedback controller gain matrix $\m K$. These types of feedback controllers are highly popular in the modern control theory---see \cite{VeeracharyITIA2022}\cite{ChengITSMC2021}. This is mainly because of their robustness toward noise/disturbance and due to the availability of computationally efficient algorithms in the literature to design them. Notice that in this design, the controller only takes the information of the state-space matrices given in model \eqref{eq:final_NODE}. To that end, $\mathcal{H}_\infty$ NODE-based controller can be designed by solving the following well-known continuous-time algebraic Riccati equation (CARE) (with a bit of abuse of notation, the dependency on $\mu$ has been removed due to space limitation):
	\begin{align*}
		\hspace{-0.08cm}\Bar{\mA}^\top\hspace{-0.01cm}\mP_d \hspace{-0.01cm} + \hspace{-0.08cm}\mP_d\Bar{\mA}\hspace{-0.01cm}+\hspace{-0.01cm}\mP_d\mG\mP_d\hspace{-0.01cm}
	\end{align*}
	\vspace{-0.3cm}
	$$ -\hspace{-0.01cm}(\mP_d\Bar{\mB}\hspace{-0.01cm}+\hspace{-0.01cm}\mS)\mR^{-1}(\Bar{\mB}^\top\hspace{-0.01cm}\mP_d\hspace{-0.01cm}+\hspace{-0.01cm}\mS^\top)\hspace{-0.03cm}+\hspace{-0.03cm}\mQ \hspace{-0.01cm}=\hspace{-0.01cm} \mO$$
	where the matrices $\Bar{\mA}, \Bar{\mB}, \m Q,\dots$ are given in Appendix~\ref{appndix:OP3}.
	We compute the $\mathcal{H}_\infty$ state-feedback controller $\mK$ as
	\begin{align}\label{eq:K_Hinf_Care}
		\mK(\mu) = \bmat{-\mR(\mu)^{-1}(\Bar{\mB}(\mu)^\top\mP_d(\mu)+\mS(\mu)^\top)&\mO}.
	\end{align}
	Utilizing the MATLAB built-in function
	\texttt{icare()} and setting  $\mE = \mI$, we can efficiently solve the above CARE. Then, utilizing the bisection method for the following optimization problem:
	\begin{align*}
		\mathbf{\left( OP_2\right) }\;\;\;\;\minimize_{{\mu}} &\;\;\; \mu\\ \subjectto  & \;\;\;\m{K}(\mu)\;\;\; \mr{in}\;\;\;\;\eqref{eq:K_Hinf_Care}\;\;\;\mr{exists}
	\end{align*} 
	we compute the optimal $\mu^*$ and $\mK(\mu^*)$. We use the structure \texttt{info.Report} of \texttt{icare()} in MATLAB to verify the existence of $\mK(\mu)$ in our bisection method. If \texttt{info.Report} $= 0$ and $\mP_d \succ \mO$, then a unique accurate $\mK(\mu)$ in \eqref{eq:K_Hinf_Care} exists. We name the controller computed from $\mathbf{OP_2}$ as $\mathcal{H}_{\infty}$-ODE. The complete derivation for this controller design is given in Appendix \ref{appndix:OP3}, which is similar to the derivation in \cite{zhou1988algebraic}.
	\subsubsection{{H-2-based ODE state-feedback controller design}}
	Similarly, for the perturbed NODE dynamics \eqref{eq:final_NODE_peturbed}, the $\mathcal{H}_2$-based state-feedback controller can also be designed as follows:
	\begin{align*}
		\mathbf{\left( OP_3\right) }\;\;\;\;\;\minimize_{{\m u_d}} &\;\;\; J(\m u_d) \\ \subjectto  & \;\;\;\mr{Dynamics}\; \;\eqref{eq:final_NODE_peturbed},\;\;\; \Tilde{\mD}_w = \m O \;
	\end{align*}
	where $J(\m u_d)$ is given as follows:
	\begin{align*}
		J(\m u_d)& = \int_0^{\infty} \m z_2^\top \m z_2 dt = \int_0^{\infty} (\tilde{\m C} \m x_d + \tilde{\m D} \m u_d)^\top (\tilde{\m C} \m x_d + \tilde{\m D} \m u_d)dt\\
		& = \int_0^{\infty} \m x_d^\top \bar{\m Q} \m x_d + \m u_d^\top \bar{\m R} \m u_d + 2 \m x_d^\top \m N \m u_d \;dt
	\end{align*}
	with
	\begin{align*}
		& \bar{\mQ} = \tilde{\mC}^\top\tilde{\mC},~\bar{\mR} = \tilde{\mD}^\top\tilde{\mD},~\mN = \tilde{\mC}^\top\tilde{\mD}.
	\end{align*}
	It is a classical result \cite{kwakernaak1969linear} that the optimal solution of $\mathbf{OP_3}$ has the form of $\m u_d = \m K_d \m x_d$ with $\m K_d = -\bar{\mR}^{-1}(\Bar{\mB}^\top\m X_d+\mN ^\top)$ where $\mX_d$ denotes the unique solution of the following CARE:
	\begin{align}\label{eq:H2_care-ode}
		\hspace{-0.2cm}\Bar{\mA}^\top\m X_d \hspace{-0.02cm} + \hspace{-0.02cm}\m X_d \Bar{\mA}-(\m X_d \Bar{\mB}+\mN)\bar{\mR}^{-1}(\Bar{\mB}^\top\m X_d+\mN^\top) \hspace{-0.01cm}+ \hspace{-0.01cm}\bar{\mQ}= \m O
	\end{align}
	where in this case the  controller gain matrix $\mK$ can be retrieved as:
	\begin{align}\label{eq:K_H2Care_ode}
		\mK = \bmat{-\bar{\mR}^{-1}(\Bar{\mB}^\top\m X_d+\mN^\top)& \mO}.
	\end{align}
	Utilizing the MATLAB built-in function
	\texttt{lqr()}, we can efficiently solve the CARE \eqref{eq:H2_care-ode}. We name this controller as $\mathcal{H}_{2}$-ODE feedback controller.
	
	%
	
	Notice that the above ODE type controllers proposed in this section are widely applied in the modern control theory---see \cite{SinghTPWRS2016, AranyaICSM2019, BazrafshanITSG2019}. However, to the best of the authors' knowledge, no such work has still been carried out to access their applicability in controlling renewable heavy interconnected power system models. Moreover, as compared to \cite{SinghTPWRS2016, AranyaICSM2019} the ODE controller designed in this work considers robust $\mathcal{H}_\infty$ or $\mathcal{H}_2$ stability notion to handle uncertainties in load demand and renewables. Furthermore as compared to \cite{  BazrafshanITSG2019} we also consider nonlinearity in the controller architecture by modeling it as an $\mathcal{L}_2$-norm-bounded disturbance. In particular, the above-formulated ODE controller architectures are unique on their own as it includes constraints and objectives that allow for the ODE controllers to be practically implemented on renewable heavy NDAE state-space representation of power systems.
	\begin{figure}[ht]
		{\includegraphics[keepaspectratio,scale=0.125]{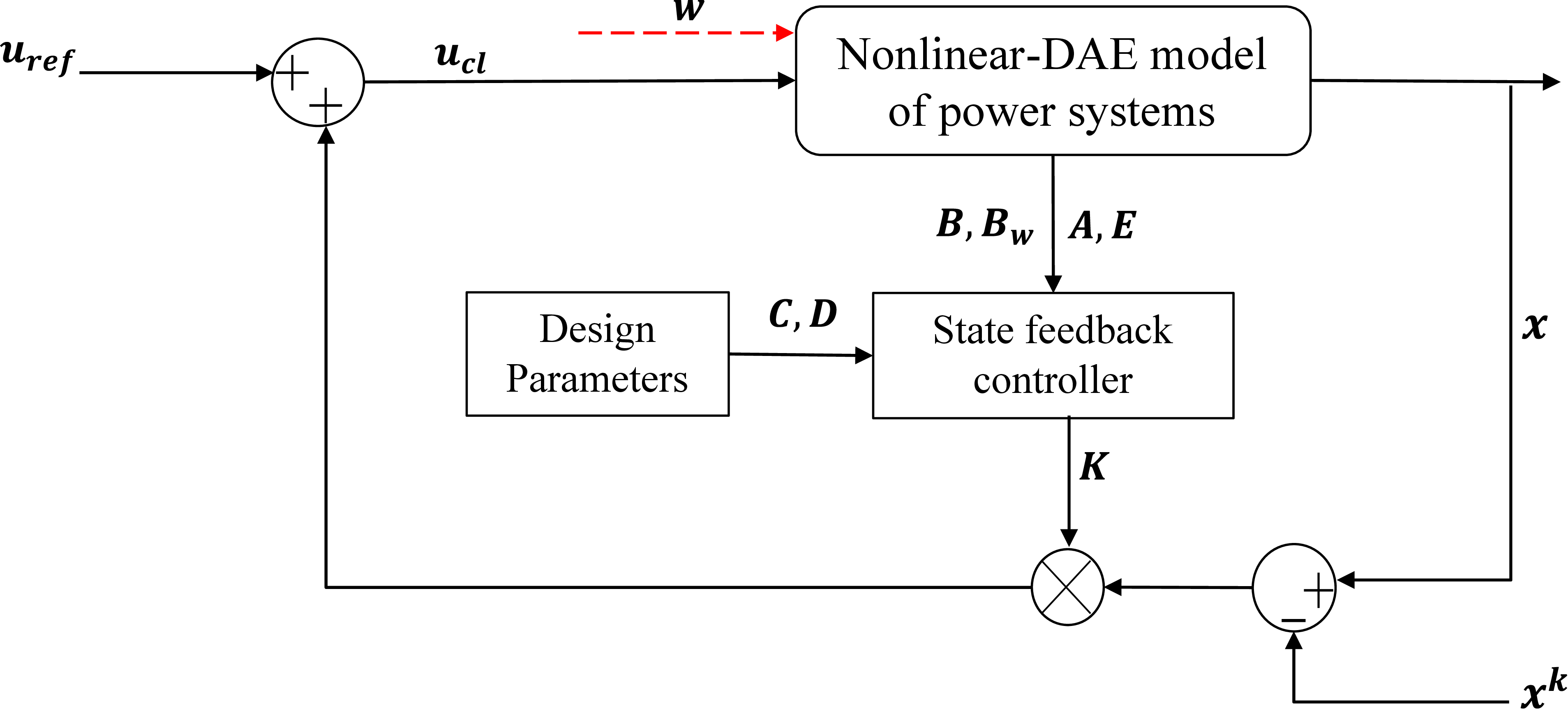}}\caption{Integrated framework of the proposed controllers. $\m x$ represents the overall state vector and encapsulates all algebraic and dynamic states while $\m w$ models load demand and solar irradiance.}\label{fig:controller design}
	\end{figure}
	\subsection{Implementation of the Proposed Feedback Controllers}
	All the proposed feedback controllers can be implemented in a similar fashion. Notice that the only major difference among them is the computation of the controller gain $\mK$. Each controller utilizes a different optimization problem to compute its corresponding $\mK$. With that in mind, the proposed feedback controllers can be implemented as follows: First, using constant system matrices given in model \eqref{eq:final_NDAE} or \eqref{eq:final_NODE} the feedback controller gain matrix $\mK$ is determined by solving the corresponding optimization problem. Then, based on the given/forecast overall load demand and solar irradiance (i.e., vector $\m w^k$) power flow is carried out, and algebraic variables of the system $\m x_a^k$ are determined. Afterward, by setting $\m E\dot{\m x} = \m0$ in \eqref{eq:final_NDAE} and using $\m x_a^k$, $\m w^k$, the reference set points for control inputs $\m u_\mr{ref}$ and steady-state values of dynamic variables $\m x_d(0)$ can be determined. Using these steady-state values (i.e., $\m x_d(0)$, $\m x_a^k$, $\m w^k$, and $\m u_\mr{ref}$) the real-time values of system state vector $\m x$ has been determined.  This state vector has then been used as state feedback in the controller design as shown in Fig. \ref{fig:controller design}. This process can be repeated for every dispatch time period.

	\section{Case Studies}\label{sec:case studies}
	
	\begin{figure}[t]
		\centering
		{\includegraphics[keepaspectratio,scale=0.13]{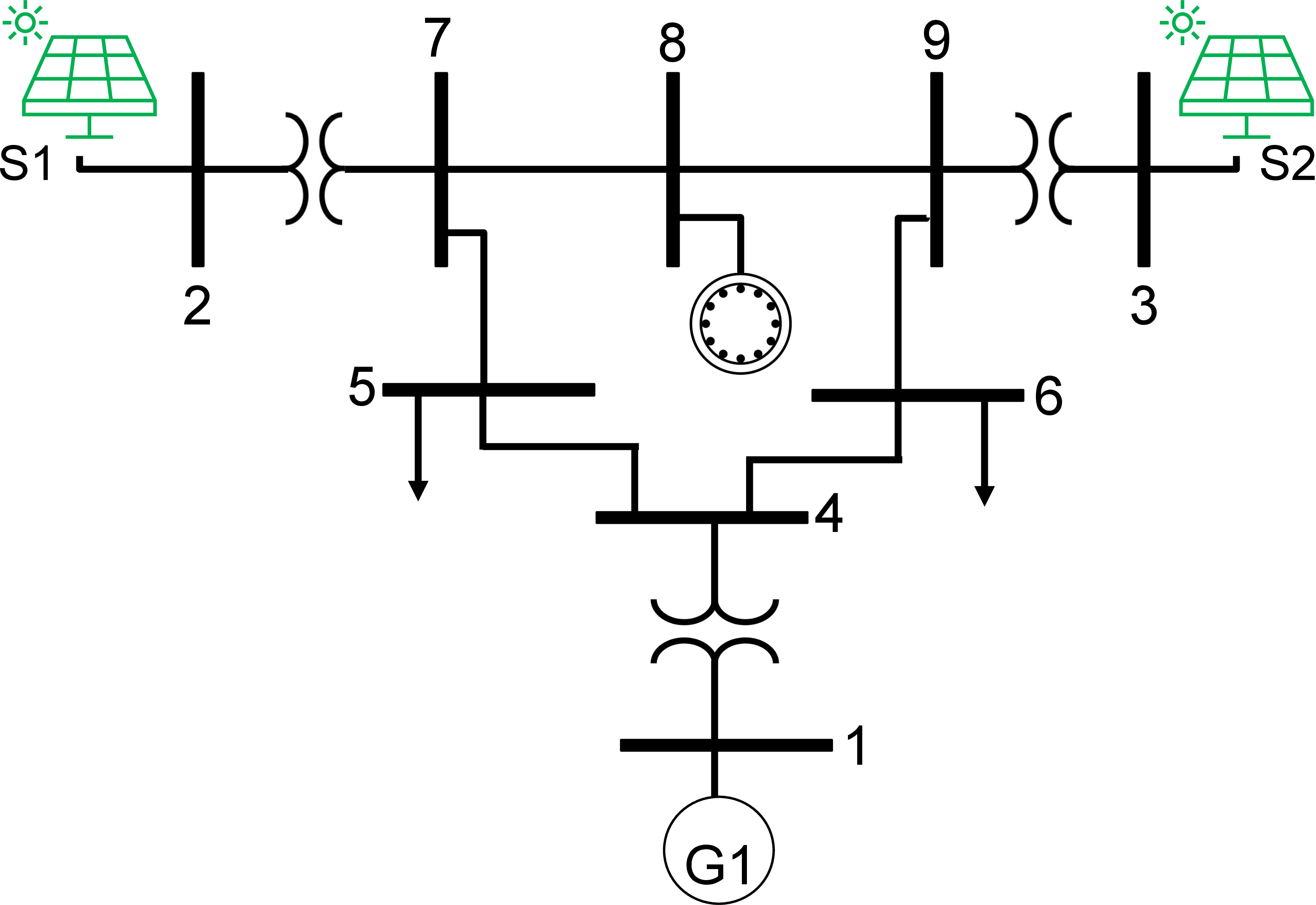}}\vspace{-0.2cm}\caption{\textcolor{black}{One line diagram of the modified WECC test power system with a motor load at Bus $8$, a synchronous generator at Bus $1$, and two solar power plants $S1$ and $S2$ at Buses $2$ and $3$, respectively. }}\label{fig:case9}\vspace{-0.3cm}
	\end{figure}
	
	\begin{figure}[t]
		\centering
		{\includegraphics[keepaspectratio,scale=0.55]{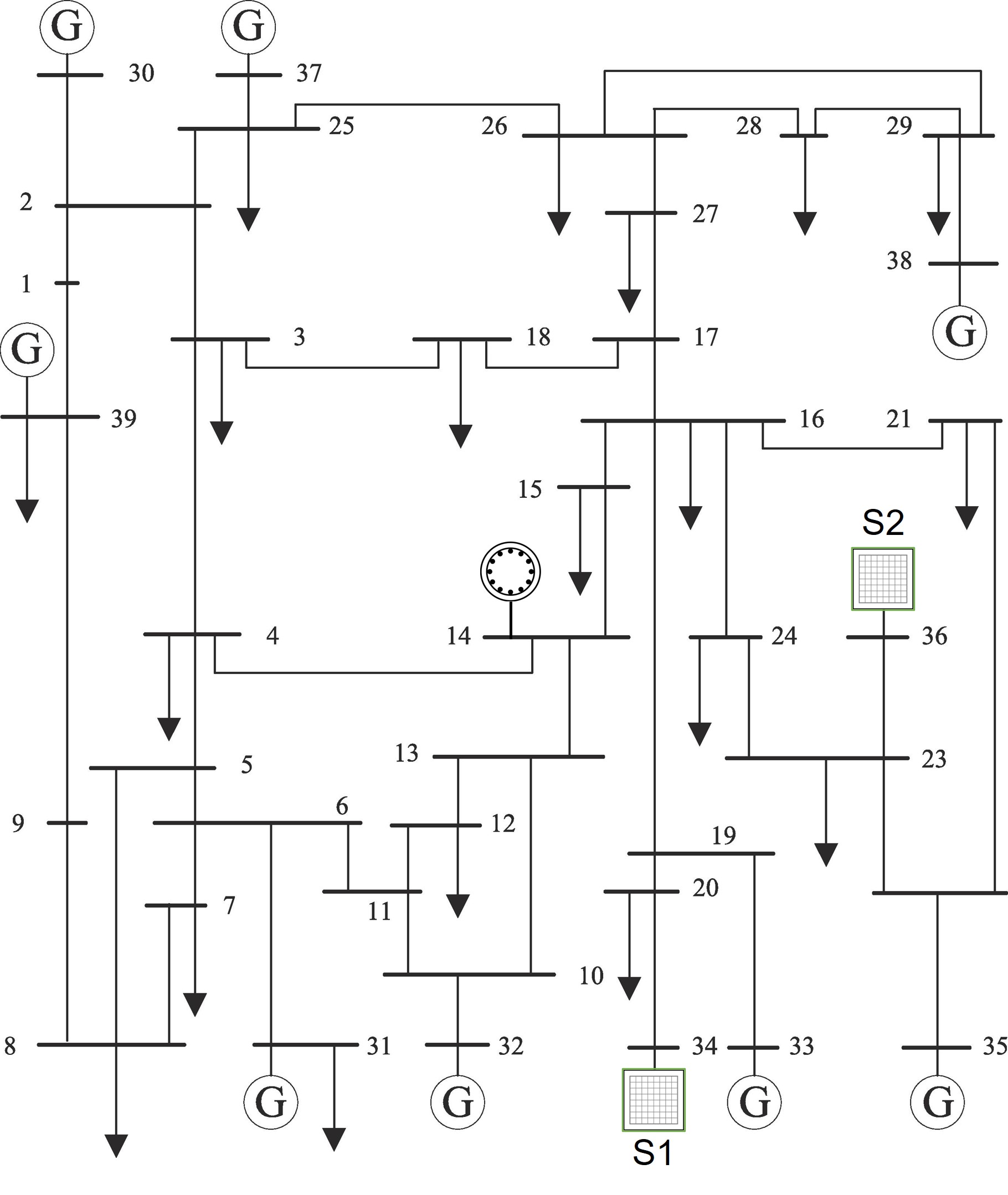}}\vspace{-0.2cm}\caption{\textcolor{black}{One line diagram of the modified IEEE $39$-bus test power system with a motor load at Bus $14$, and two solar power plants $S1$ and $S2$ at Buses $34$ and $36$, respectively.} }\label{fig:case39}\vspace{-0.6cm}
	\end{figure}
	
	Here, we demonstrate the effectiveness of the proposed feedback controllers in improving the transient stability of an interconnected power system model after a large disturbance. In specific, we try to answer the following questions:
	\begin{itemize}
		
		%
		
		\item [--] \textit{Q1:} How do the proposed controllers affect the dynamic response of an interconnected power system model after a large disturbance? How is the system's transient stability with and without the proposed feedback controllers?
		\item [--] \textit{Q2:} Does the feedback controller based on the knowledge of the complete NDAE model outperform the NODE model counterpart in terms of damping system oscillations? How important is the knowledge of dynamic and algebraic states in the feedback control of power systems?
		\item [--] \textit{Q3:} How much are NDAE and NODE-based feedback controllers robust against unknown faults/disturbances in load demand and renewables?
		\item [--] \textit{Q4:} How computationally scalable are these feedback controllers when applied to a larger power system model?
	\end{itemize}
	The proposed feedback controllers have been tested on modified IEEE $9$-bus and $39$-bus power systems \cite{Anderson2003,Hiskens}. \textcolor{black}{The one-line diagrams of these test systems are shown in Figs. \ref{fig:case9} and \ref{fig:case39}}. All the parameters for the motor load and synchronous generators along with its excitation system can be found in \cite{sauer2017power, krause2013}, while the detailed description of the solar farm model and its parameters can be seen in \cite{SoumyaITPWRS2022, Hart2016}.
	
	The numerical case studies have been carried out in MATLAB R$2022a$ running on a PC with an Intel $i9-11980$HK processor and $64$GB of RAM. The  power system model \eqref{eq:final_NDAE} is solved using MATLAB's DAEs system solver \texttt{ode15s} with settings chosen to be: (i) maximum step size $=1\times10^{-5}$ (ii) absolute tolerance $=1\times10^{-7}$, and (iii) relative tolerance $=1\times10^{-7}$. The optimization problem $\mathbf{OP_1}$ is solved in YALMIP \cite{LofbergICRA2004} using MOSEK solver \cite{Andersen2000}. The power system volt-ampere base is considered as $S_b = 100\mr{MVA}$ while the frequency base is chosen to be $w_b = 120\pi\mr{rad/s}$.
	The initial conditions and steady-state values of the power system before any disturbance/fault are determined using power flow studies carried out in MATPOWER \cite{Matpower} via function \texttt{runpf}. \textcolor{black}{ Furthermore, in all the case studies we assume that all states of the power system are available in realtime. This is reasonable as there exist efficient observers that can estimate all the states including the states of solar plants and motor loads via few PMUs optimally placed in the network as detailed in \cite{nadeem2022robust}}.
	
	
	\vspace{-0.2cm}
	\subsection{Performance under Large Disturbance in Load Demand}
	\begin{figure}[ht]
		\subfloat{\includegraphics[keepaspectratio=true,scale=0.53]{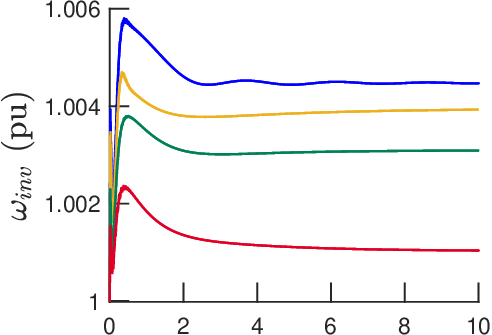}}{}\hspace{-0.05cm}
		\subfloat{\includegraphics[keepaspectratio=true,scale=0.53]{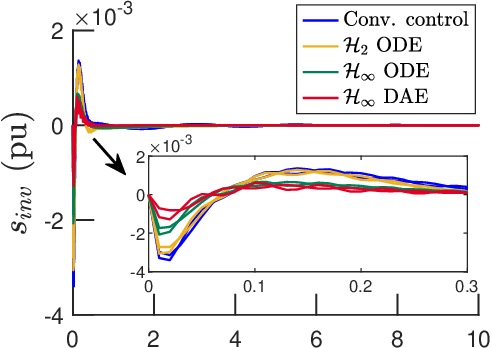}}{}{}\hspace{-0.25cm}
		\subfloat{\includegraphics[keepaspectratio=true,scale=0.53]{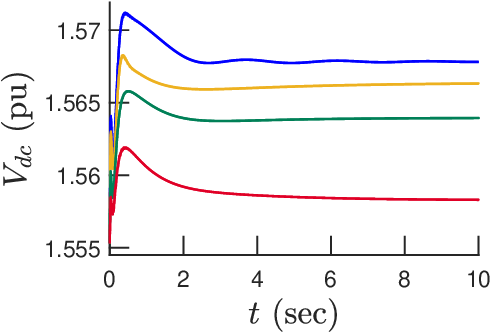}}{}{}\vspace{0.19cm}\hspace{-0.05cm}
		\subfloat{\includegraphics[keepaspectratio=true,scale=0.53]{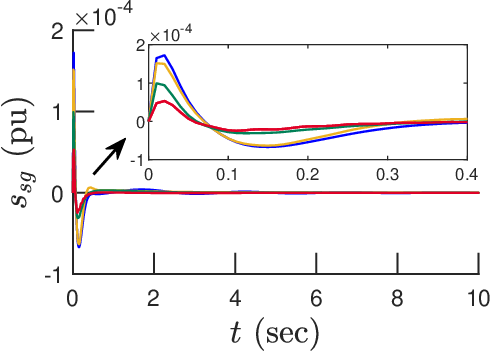
		}}{}{}\vspace{-0.5cm} \caption{Performance under sudden decrease in load demand for IEEE $9$-bus system: angular speed of both solar plants $S1$ and $S2$, both inverters relative slip,  DC link voltage, and generator slip.}\label{fig:pd decrease case 9}
		\vspace{-0.4cm}
	\end{figure}
	
	\begin{figure}[ht]
		\subfloat{\includegraphics[keepaspectratio=true,scale=0.53]{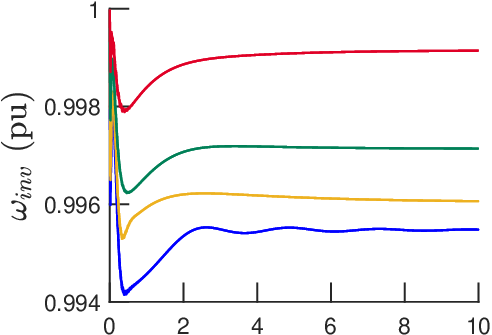}}{}\hspace{-0.05cm}
		\subfloat{\includegraphics[keepaspectratio=true,scale=0.53]{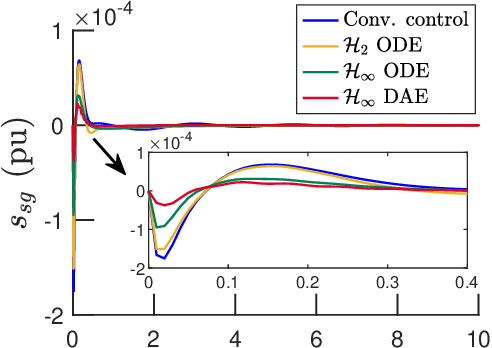}}{}{}\hspace{-0.25cm}
		\subfloat{\includegraphics[keepaspectratio=true,scale=0.53]{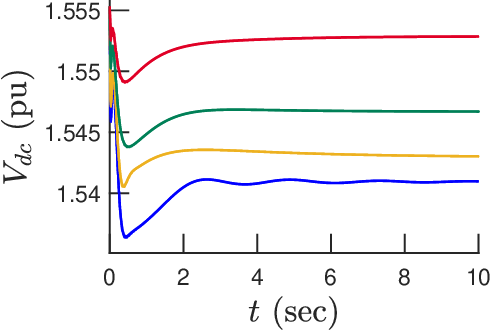
		}}{}{}\hspace{-0.05cm}\vspace{-0.5cm}
		\subfloat{\includegraphics[keepaspectratio=true,scale=0.53]{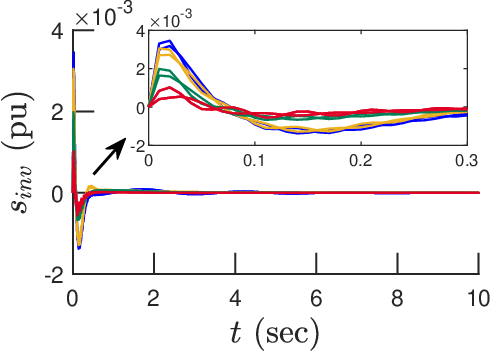}}{}{}\vspace{0.19cm}
		\caption{Performance under a sudden increase in load demand for IEEE $9$-bus system: inverters relative speed for both $S1$ and $S2$, generator slip, DC link voltage, and relative slip for both inverters.}\label{fig:pd increase case 9}
		\vspace{-0.4cm}
	\end{figure}
	
	
	\begin{figure}[ht]
		\subfloat{\includegraphics[keepaspectratio=true,scale=0.53]{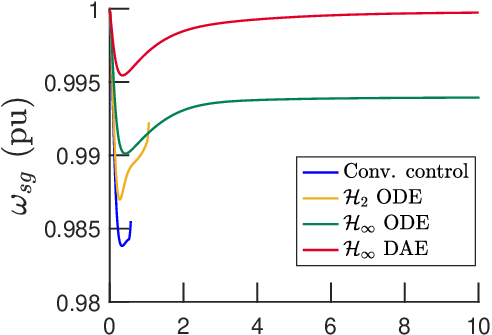}}{}\hspace{-0.05cm}
		\subfloat{\includegraphics[keepaspectratio=true,scale=0.53]{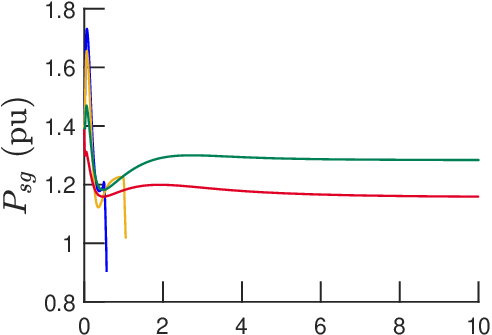}}{}{}\vspace{0.19cm}\hspace{-0.05cm}
		\subfloat{\includegraphics[keepaspectratio=true,scale=0.53]{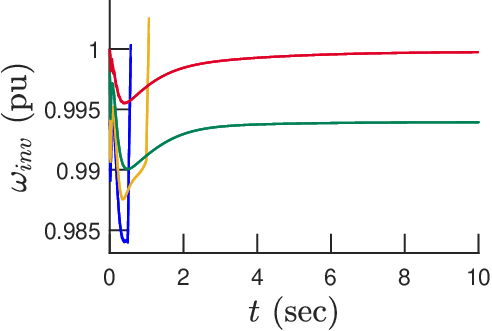}}{}{}\hspace{-0.05cm}
		\subfloat{\includegraphics[keepaspectratio=true,scale=0.53]{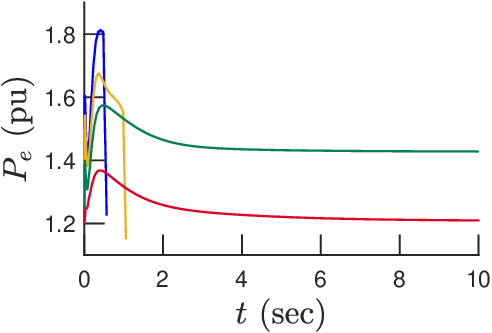
		}}{}{}\vspace{-0.5cm} \caption{Performance under $\Delta_d = 0.5$ and $\Delta_I = 0.2$ for IEEE $9$-bus system:  generator speed,  power supplied by a synchronous generator, both inverters relative speed, and power supplied by PV plant $S1$.}\label{fig:case 9 delta-d 1.8 and delta-1 0.2}
		\vspace{-0.4cm}
	\end{figure}
	
	\begin{figure}[ht]
		\subfloat{\includegraphics[keepaspectratio=true,scale=0.53]{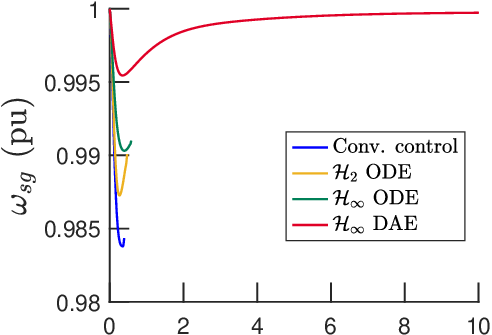}}{}\hspace{-0.05cm}
		\subfloat{\includegraphics[keepaspectratio=true,scale=0.53]{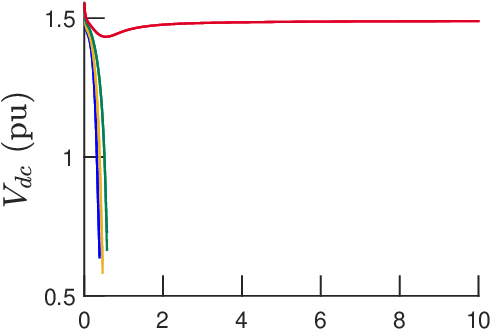}}{}{}\hspace{-0.25cm}
		\subfloat{\includegraphics[keepaspectratio=true,scale=0.53]{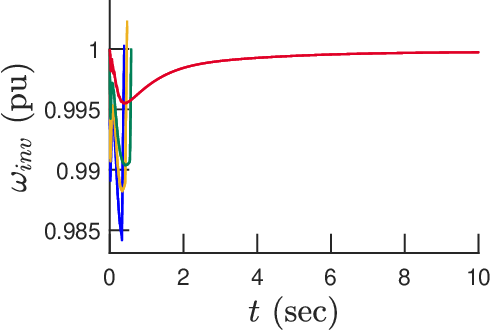}}{}{}\vspace{0.19cm}\hspace{-0.05cm}
		\subfloat{\includegraphics[keepaspectratio=true,scale=0.53]{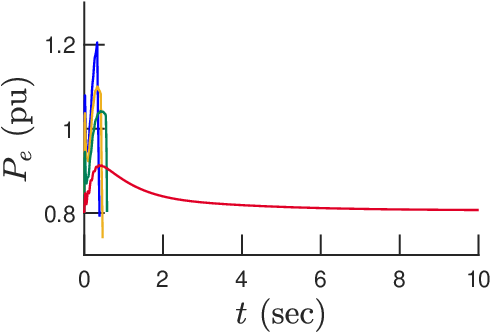
		}}{}{}\vspace{-0.5cm} \caption{Performance under $\Delta_d = 0.6$ and $\Delta_I = 0.3$ for IEEE $9$-bus system:  generator speed, DC link voltage, both inverters relative speed, and power supplied by PV plant $S2$.}\label{fig:case 9 delta-d 2.5 and delta-I 0.3}
		\vspace{-0.4cm}
	\end{figure}
	
	\begin{figure}[ht]
		\subfloat{\includegraphics[keepaspectratio=true,scale=0.53]{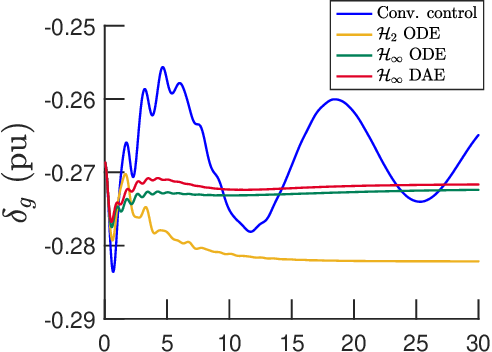}}{}\hspace{-0.05cm}
		\subfloat{\includegraphics[keepaspectratio=true,scale=0.53]{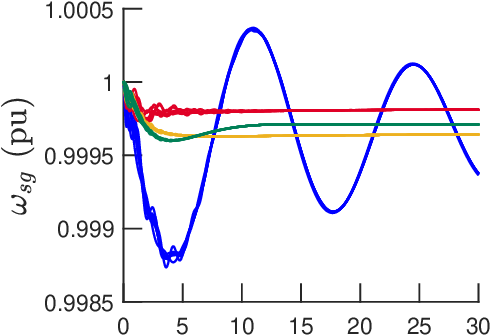}}{}{}\hspace{-0.25cm}
		\subfloat{\includegraphics[keepaspectratio=true,scale=0.53]{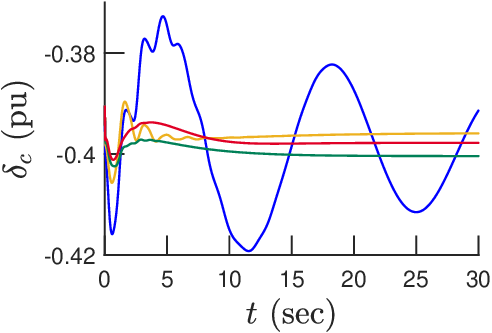}}{}{}\vspace{0.19cm}\hspace{-0.05cm}
		\subfloat{\includegraphics[keepaspectratio=true,scale=0.53]{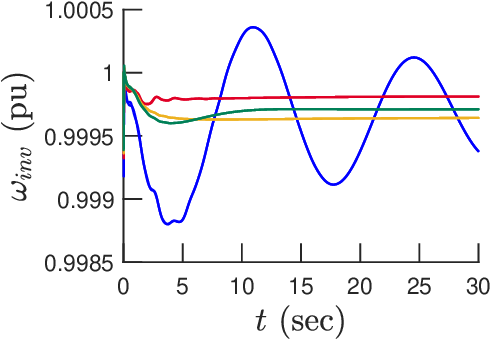
		}}{}{}\vspace{-0.5cm} \caption{Performance under disturbance in load demand and sun irradiance for IEEE $39$-bus system:  Generator $2$ rotor angle, all generators speed, $S2$ relative angle, and relative speed of both $S1$ and $S2$.}\label{fig:case 39 first case}
	\end{figure}
	
	\begin{figure}[ht]
		\subfloat{\includegraphics[keepaspectratio=true,scale=0.53]{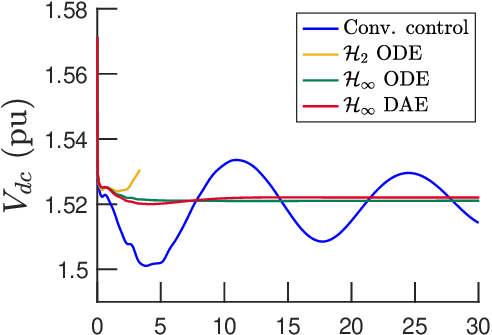}}{}\hspace{-0.05cm}
		\subfloat{\includegraphics[keepaspectratio=true,scale=0.53]{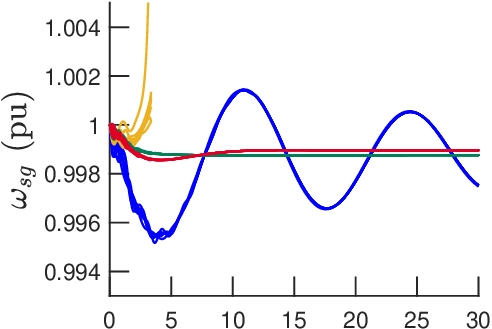}}{}{}\hspace{-0.25cm}
		\subfloat{\includegraphics[keepaspectratio=true,scale=0.53]{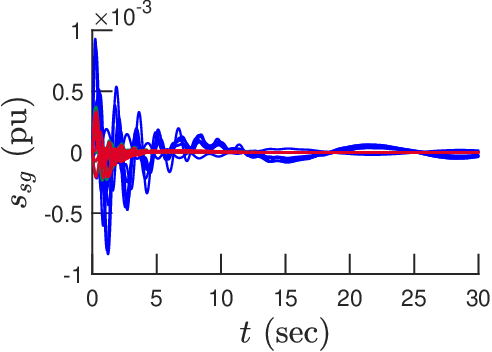}}{}{}\vspace{0.19cm}\hspace{-0.05cm}
		\subfloat{\includegraphics[keepaspectratio=true,scale=0.53]{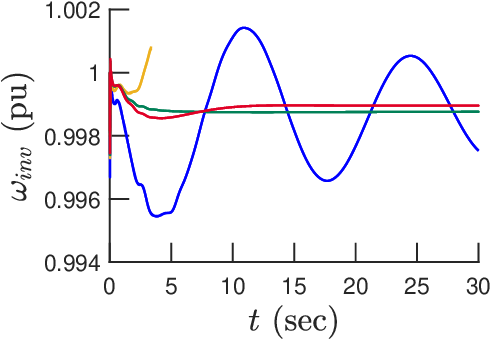
		}}{}{}\vspace{-0.5cm} \caption{Performance under $\Delta_d = 0.01$ and $\Delta_I = 0.3$ for IEEE $39$-bus system: $S2$ DC link voltage, generators speed, generators slip, and relative speed of both $S1$ and $S2$.}\label{fig:case 39 delta-d 0.01 delta-I 0.3}
	\end{figure}
	
	\begin{figure}[ht]
		\subfloat{\includegraphics[keepaspectratio=true,scale=0.53]{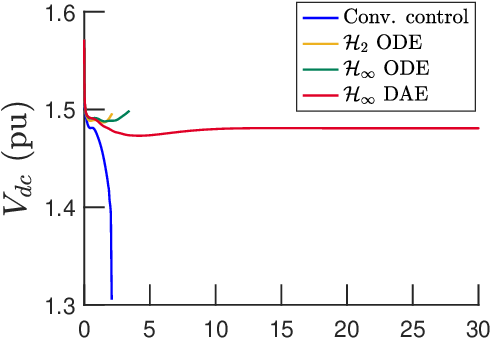}}{}\hspace{-0.05cm}
		\subfloat{\includegraphics[keepaspectratio=true,scale=0.53]{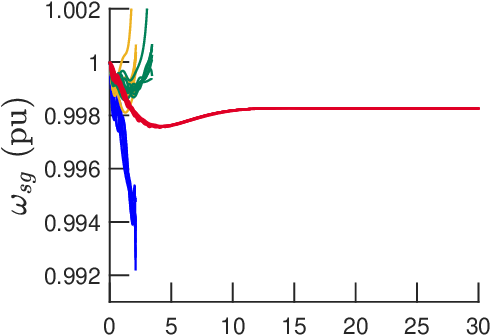}}{}{}\hspace{-0.25cm}
		\subfloat{\includegraphics[keepaspectratio=true,scale=0.53]{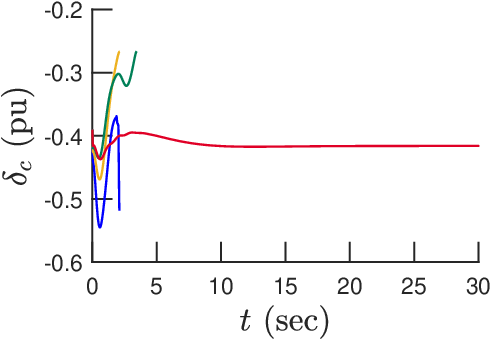}}{}{}\vspace{0.19cm}\hspace{-0.05cm}
		\subfloat{\includegraphics[keepaspectratio=true,scale=0.53]{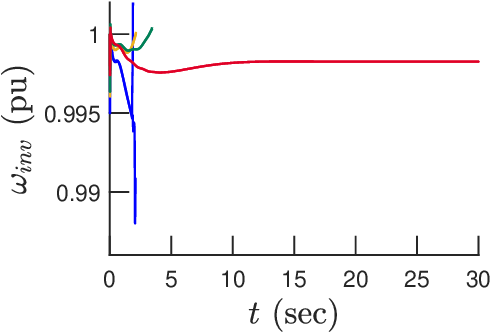
		}}{}{}\vspace{-0.5cm} \caption{Performance under $\Delta_d = 0.03$ and $\Delta_I = 0.4$ for IEEE $39$-bus system: $S2$ DC link voltage, generators speed, Generator $2$ rotor angle, and relative speed of both $S1$ and $S2$.}\label{fig:case 39 worst scenior}
		\vspace{-0.4cm}
	\end{figure}
	
	\begin{table}[t]
		\caption{Comparison of Computational time for calculating gain matrix using MOSEK as optimization solver}
		\label{tab:Table 1}
		\centering
		\begin{tabular}{c|ccc}
			\hline
			\begin{tabular}[c]{@{}c@{}}Test \\ System\end{tabular} &
			\multicolumn{3}{c}{\begin{tabular}[c]{@{}c@{}}Computational time for calculating gain matrix \\ $\mK$, $t (s)$\end{tabular}} \\ \hline
			& \multicolumn{1}{c|}{$\mathcal{H}_{\infty}$ ODE} & \multicolumn{1}{c|}{$\mathcal{H}_{2}$ ODE} & $\mathcal{H}_{\infty}$ DAE \\ \hline
			IEEE $9$-bus system  & \multicolumn{1}{c|}{0.589}                      & \multicolumn{1}{c|}{0.152}                 & 4.567                      \\ \hline
			IEEE $39$-bus system & \multicolumn{1}{c|}{0.671}                      & \multicolumn{1}{c|}{0.356}                 & 14760                      \\ \hline
		\end{tabular}
	\end{table}
	
	
	In this section, we discuss the performance of the proposed controller under large-step disturbance in the overall load demand of the power system. With that in mind, the simulations are carried out as follows: 
	In the beginning, the system operates with the overall load demand of $P^0_d + Q^0_d =0.77 + j0.25 $ $\mr{pu}$ for Case $9$-bus system and $P^0_d + Q^0_d =19.8 + j7.10 $ $\mr{pu}$ for the $39$-bus system. The total power generation (from both renewables and conventional power plants) is equal to the load demand and thus there are no transients in the power system and all the states of the network rest at their equilibrium values. Then, suddenly after $t>0$, the load demand is abruptly changed and their new value is given as $P^e_d + Q^e_d =(1+\Delta_d)(P^0_d + Q^0_d)$, where $\Delta_d$ represent the amount of the disturbance. In this work, two simulations studies have been carried out for Case $9$-bus system with $\Delta_d$ chosen to be $-0.4$ and $0.4$ while for Case $39$-bus system we select  $\Delta_d = 0.001$. Notice that selecting  $\Delta_d$ to be negative means that the overall load demand has been decreased (or a load trip event occurred in the system) while choosing $\Delta_d$ positive shows an abrupt increase in the system load demand (which can roughly be presumed as a generator trip event in the power network). To mimic realistic changes in load demand a Gaussian noise $q_d(t)$ with zero mean and variance of $0.01\Delta_d$ has also been added and thus the overall system load disturbance can be written as  $P^e_d + Q^e_d =(1+\Delta_d)(P^0_d + Q^0_d) + q_d(t)$.
	
	When this disturbance is applied, it will trigger the power system to depart from its steady-state conditions and push it to a new equilibrium or even possibly make it lose synchrony. The main objective of the controller is to minimize the impact of the disturbance $\Delta_d$ on the system dynamics and thus provide damping to the system oscillations during transient conditions and restore the system back to its nominal value as soon as possible. 
	
	The results of these simulation studies are presented in Figs. \ref{fig:pd decrease case 9},  \ref{fig:pd increase case 9}, and \ref{fig:case 39 first case}. To showcase the performance of the proposed controllers a comparison between the system response after a large disturbance with only conventional control (or primary control) and with proposed feedback controllers acting on top of them has also been presented. Notice that, by conventional control, we refer to the legacy primary controllers of the power system, which for the synchronous generator in this test system consists of AVRs, PSSs, and governors. Similarly, solar farms are acting in grid-forming mode and their primary controller consists of droop and proportional-integral (PI) type controllers. Further details about these conventional controllers can be found in \cite{sauer2017power, DudgeonITPWRS2007} for synchronous generators and in \cite{SoumyaITPWRS2022, WasynczukITPE1996} for solar farms. Notice that these conventional controllers are already present in the test system and the proposed feedback controllers are acting on top of them and are sending additional control signal $\m u_\mr{wac}$ to effectively mitigate the effect of disturbance $\Delta_d$ on power system dynamics.
	
	
	We plot inverter frequency $\m\omega_\mr{inv}$ and slip $\m s_\mr{inv}$ in Figs. \ref{fig:pd decrease case 9} and  \ref{fig:pd increase case 9} which are computed from state vector as follows:
	\begin{align*}
		\m\omega_\mr{inv} &=  1-k_\mr{inv}(\tilde{\mP_e}-\mP_s^*), \; \; \m s_\mr{inv} = (w_e - \m\omega_\mr{inv})/w_e
	\end{align*}
	where $w_e$ denotes the center of inertia mean angular speed of the overall power system \cite{SoumyaITPWRS2022}, $\tilde{\mP_e}$ is the phasor representations of ${\m P}_{e}$ (active power outflows of the inverters to the grid) after passing through low pass filter, and $k_\mr{inv}$ is the droop constant of the inverter. From Figs. \ref{fig:pd decrease case 9}, \ref{fig:pd increase case 9}, and \ref{fig:case 39 first case}  we can clearly see that with the proposed robust wide-area feedback controllers there is significant damping in frequency oscillations during the transient period (first few seconds after the disturbance) and also the frequency nadir has been significantly improved. For the Case $39$-bus test system we can see from Fig. \ref{fig:case 39 first case} that after a large disturbance in load demand the generator's angular frequency dips to near $0.998$ $\mr{pu}$ and starts oscillating, while with robust WACs as a secondary control loop, the generator frequency decrease very slightly and the frequency oscillations have significantly been damped out.

	
	Similar results have also been achieved for all synchronous generators and inverter relative speeds for the Case $9$-bus test system as shown in Figs. \ref{fig:pd decrease case 9} and \ref{fig:pd increase case 9}. We can clearly see that with the proposed robust WACs, there is a significant damping in the system oscillations and the RoCoF is very less, thus the overall transient stability of the power system after a large disturbance has been improved.
	
\begin{figure}[ht]
		\centering
		\subfloat{\includegraphics[keepaspectratio=true,scale=0.53]{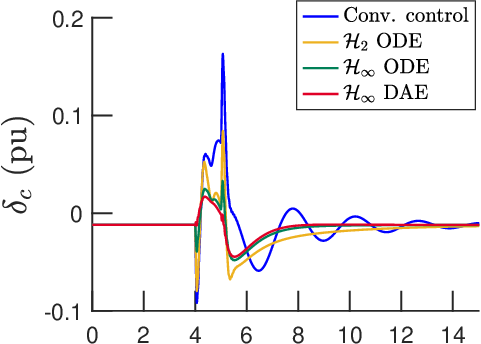}}{}\hspace{0.05cm}
		\subfloat{\includegraphics[keepaspectratio=true,scale=0.53]{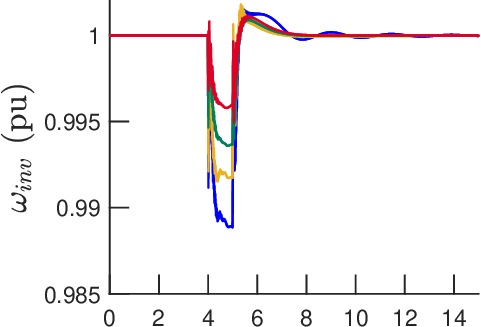}}{}\hspace{-0.05cm}
		\subfloat{\includegraphics[keepaspectratio=true,scale=0.53]{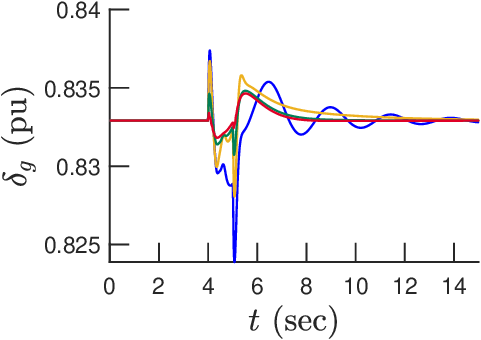}}{}\hspace{0.05cm}
		\subfloat{\includegraphics[keepaspectratio=true,scale=0.53]{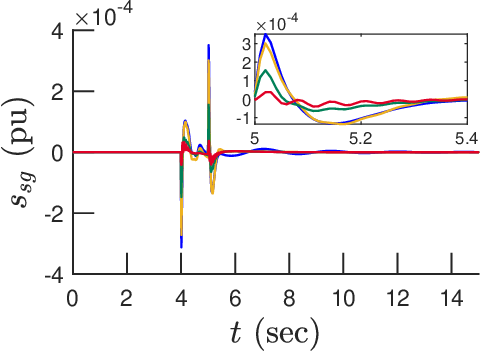}}{}\hspace{0.05cm}
		
		\vspace{-0.25cm}
		\caption{PV power plant 1 relative angle, speed, rotor angle, and rotor slip under line to ground fault, IEEE $9$-bus system.}\label{fig:fault_gen}\vspace{-0.4cm}
	\end{figure}

	To further advocate for the benefits of the proposed WACs, we also assess their performance under a short circuit fault. To that end, we added a line to ground fault at $t = 4$sec 
	on transmission line $4$-$6$ which is then cleared at $50$ msec and $200$ msec from the near and remote end for the $9$-bus test system. The simulation results are presented in Fig. \ref{fig:fault_gen}. Again, we observe that with only conventional control, the frequency nadir during the fault is higher and there are significant system oscillations thereafter. While with the proposed WACs on top of them, the overall frequency dip has been improved and there are damping in the system oscillations (particularly with $\mathcal{H}_\infty$-DAE and $\mathcal{H}_\infty$-ODE controllers), thus, improving the system transient stability.
	
	We can also see from these figures that with the proposed robust WACs, the system quickly restores to its nominal value (specifically with $\mathcal{H}_\infty$-DAE) after a large disturbance while with only conventional/primary control, the system states do not restore to its pre-fault equilibrium values and settle to new steady-state values. This also indicates that with the proposed robust WACs, the work required by the AGC, later on, to restore the system frequency to its nominal value has also been reduced.
	
	\subsection{Comparative Analysis of the Three Controllers}
	Since we have proposed three different types of wide-area feedback controllers, then regarding comparison among them, we notice that the $\mathcal{H}_\infty$-DAE based controller is providing the most damping (which can be verified from the plots of slips of generators and inverters in Figs. \ref{fig:pd decrease case 9} and \ref{fig:pd increase case 9}) then $\mathcal{H}_\infty$-ODE and finally $\mathcal{H}_2$-ODE based controller performed the worst.
	
	It is noteworthy that from a control theoretic perspective, it is well-known that $\mathcal{H}_\infty$ based controllers are more robust than $\mathcal{H}_2$ based controllers \cite{YuFend} because, in $\mathcal{H}_\infty$ stability notion, we explicitly model the disturbance vector and the controller makes sure that the strict $\mathcal{H}_\infty$ stability criterion (as explained in Def. \ref{def:H_inf}) is always satisfied. Furthermore, the reason why  $\mathcal{H}_\infty$-DAE controller is adding more damping as compared to the other feedback controllers is that it is based on a complete DAE system model and thus it sends control signals based on the knowledge of both dynamic $\m x_d$ and algebraic states $\m x_a$ as compared to the ODE based controllers which only take $\m x_d$ as state feedback. Thus the DAE-based controller has more knowledge of the system and can send more accurate control signals to the system as compared to the ODE-based controllers.
	
	However, from simulations studies, we observe that although DAE-based controllers perform better than ODE-based ones they are difficult to be scaled to larger power system models. This has been shown in Tab. \ref{tab:Table 1}, for Case $9$-bus test system we can see that the solver took around $4.56s$ to solve $\mathbf{OP_1}$ and to compute controller gain matrix $\m K$ while for Case $39$-bus the time for computing $\m K$ has been increased to around $4.1hrs$. Thus solving $\mathbf{OP_1}$ for a larger power system model with hundreds or thousands of buses can be problematic and not scalable. On the other hand for both the ODE-based controllers the computation time is less than $1s$ for all the case studies.
	
	\begin{figure}[ht]
		\centering
		\subfloat{\includegraphics[keepaspectratio=true,scale=0.53]{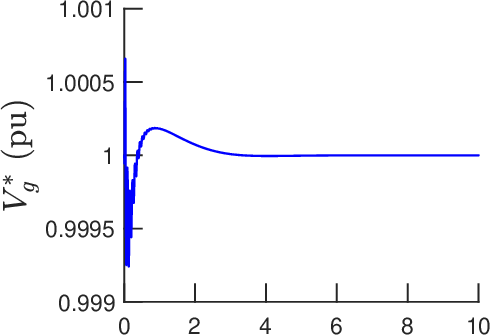}}{}\hspace{-0.10cm}
		\subfloat{\includegraphics[keepaspectratio=true,scale=0.53]{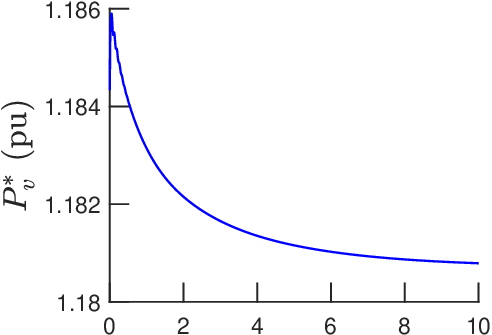}}{}\hspace{0.05cm}
		\subfloat{\includegraphics[keepaspectratio=true,scale=0.53]{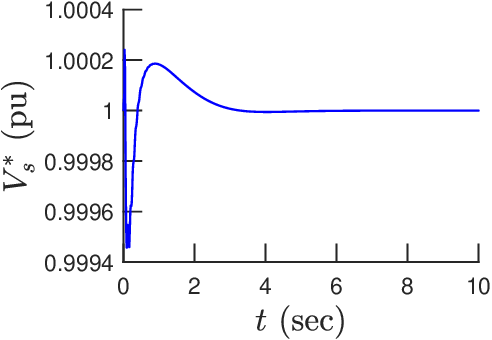}}{}\hspace{-0.05cm}
		\subfloat{\includegraphics[keepaspectratio=true,scale=0.53]{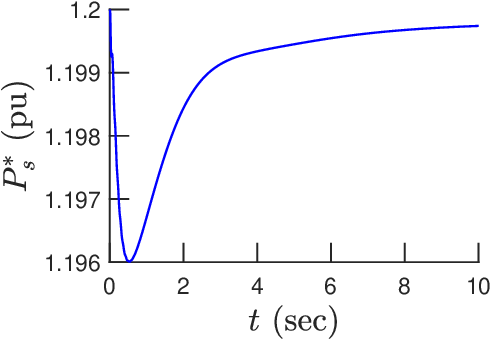}}{}\hspace{-0.25cm}
		\caption{Control inputs generated via $\mathcal{H}_\infty$-DAE controller for synchronous generator and PV power plant 1 under case C for the $9$-bus test system.}\label{fig:control input 9}
		\vspace{-0.4cm}
	\end{figure}
	
	\begin{figure}[ht]
		\centering
		\subfloat{\includegraphics[keepaspectratio=true,scale=0.53]{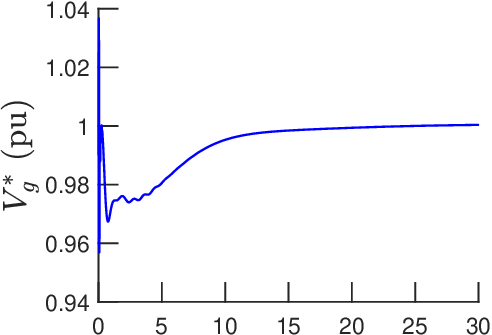}}{}\hspace{-0.10cm}
		\subfloat{\includegraphics[keepaspectratio=true,scale=0.53]{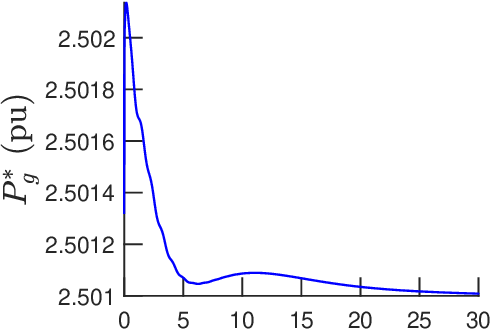}}{}
		\subfloat{\includegraphics[keepaspectratio=true,scale=0.53]{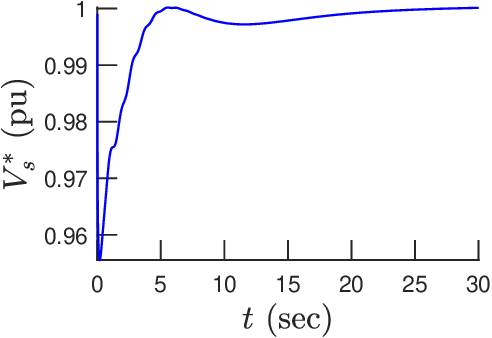}}{}\hspace{-0.10cm}
		\subfloat{\includegraphics[keepaspectratio=true,scale=0.53]{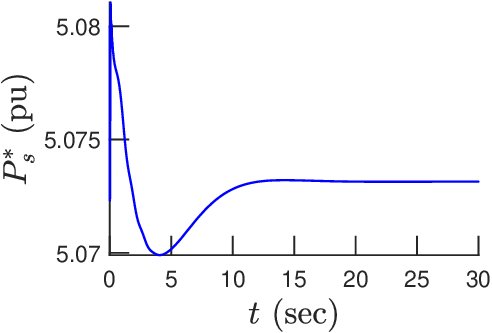}}{}\hspace{-0.25cm}
		\caption{Control inputs generated via $\mathcal{H}_\infty$-DAE controller for generator $1$ and PV power plant $1$ under case C for the $39$-bus test system.}\label{fig:control input 39}
	\end{figure}
	\begin{figure}[ht]
		\centering
		\hspace{-0.15cm}\subfloat{\includegraphics[keepaspectratio=true,scale=0.5]{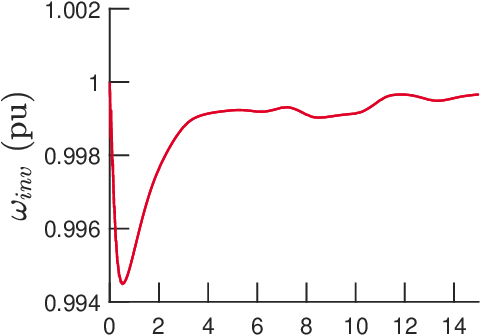}}{}\hspace{0.25cm}
		\subfloat{\includegraphics[keepaspectratio=true,scale=0.5]{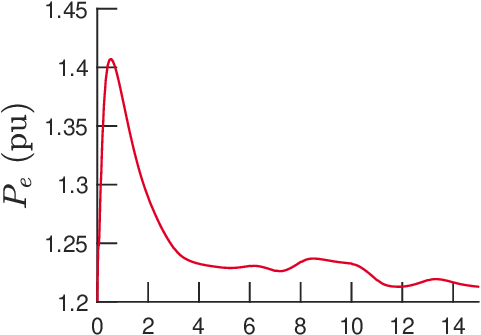}}{}\hspace{-0.05cm}	
		\subfloat{\includegraphics[keepaspectratio=true,scale=0.5]{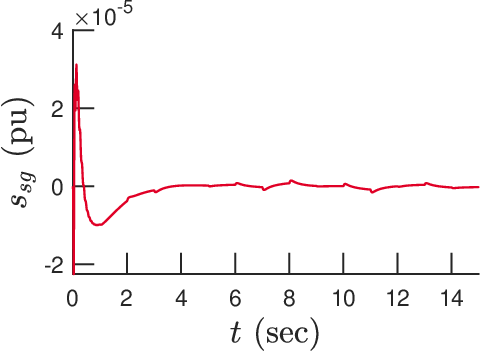}}{}\hspace{-0.05cm}
		\subfloat{\includegraphics[keepaspectratio=true,scale=0.5]{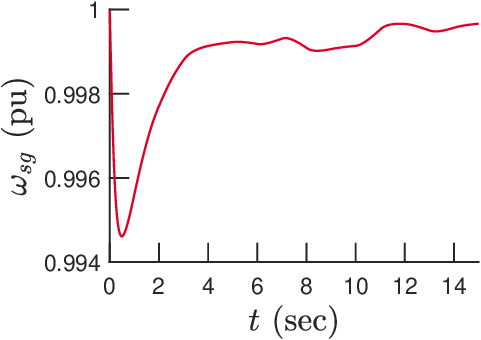}}{}\hspace{-0.05cm}\vspace{-0.25cm}
		\caption{PV plant 1 speed, PV plant 1 power, rotor speed, and rotor slip with $\mathcal{H}_\infty$-DAE controller under $\Delta_d = 0.6$, $\Delta_I = 0.3$ and noisy states measurements, IEEE $9$-bus test system. Notice that the rest of the controllers were unable to stabilize the system.}\label{fig:noise_state}
		\vspace{-0.4cm}
	\end{figure}
	\vspace{-0.5cm}
	%
	
	\subsection{Performance under Uncertainty in Irradiance and Load Demand}
	To further advocate the advantages of the proposed robust WACs in improving the transient stability of the grid and making it more robust, we add more disturbance in the system this time by changing the sun's irradiance on both PV power plants. Notice that the disturbance in load demand from the previous section has also been applied simultaneously and their severity has been further increased. To that end the simulations for this section have been carried out as follows: Initially, both PV power plants $S1$ and $S2$ operates under standard solar irradiance of $1000$ $W/m^2$ then right after $t>0$ the irradiance on both PV plants has been decreased and their new values are given as  $I^e_r =(1-\Delta_I)(I^0_r) + q_I(t)$, where $\Delta_I$ denotes the severity of the disturbance, $q_I(t)$ is Gaussian noise with zero mean and variance of $0.01\Delta_I$, and $I^0_r$, $I^e_r$ represent the sun irradiance before and after the disturbance, respectively. The disturbance in the load demand has been applied similarly as discussed in the previous section.
	
	With that in mind, numerical studies for different severity of disturbances in load demand and sun irradiance have been simulated for Case $9$-bus and Case $39$-bus test systems. For the Case $9$-bus system, we ran simulations studies for $\Delta_d = 0.5$, $\Delta_I = 0.2$ and $\Delta_d = 0.6$, $\Delta_I = 0.3$, similarly for Case $39$-bus test system we ran two simulations one with $\Delta_d = 0.01$, $\Delta_I = 0.3$ and for the other one we selected $\Delta_d = 0.03$, $\Delta_I = 0.4$. The results are presented in Figs. \ref{fig:case 9 delta-d 1.8 and delta-1 0.2}, \ref{fig:case 9 delta-d 2.5 and delta-I 0.3}, \ref{fig:case 39 delta-d 0.01 delta-I 0.3}, and \ref{fig:case 39 worst scenior}. For the Case $9$-bus system from Fig. \ref{fig:case 9 delta-d 1.8 and delta-1 0.2}, we observe that as the severity of disturbances has been increased, with only conventional/primary controllers, the system loses its synchrony and become unstable. Furthermore, we also observe that $\mathcal{H}_{2}$-ODE-based controller is also unable to stabilize the system while the  $\mathcal{H}_{\infty}$-ODE and $\mathcal{H}_{\infty}$-DAE based controllers can still stabilize the system. Similarly, from Fig. \ref{fig:case 9 delta-d 2.5 and delta-I 0.3}, we observe that as the severity of disturbances has been further increased, all the controllers are unable to stabilize the system except $\mathcal{H}_{\infty}$-DAE. This can also be corroborated by Figs. \ref{fig:control input 9} and \ref{fig:control input 39} where we present the control signals sent by
	$\mathcal{H}_{\infty}$-DAE controller to generator $1$ and PV power plant $1$ for both the $9$-bus and $39$-bus systems. We can verify that the proposed $\mathcal{H}_{\infty}$-DAE-based WAC can still send control signals during transient conditions and thus stabilize the systems. 
	
	To further assess the robustness of $\mathcal{H}_{\infty}$-DAE controller, we also add Gaussian noise with zero mean and variance of $0.01$ to the state measurements received by the WAC. We can see that $\mathcal{H}_{\infty}$-DAE can still keep the system synchronized. Notice that this is again as discussed in the previous section mainly because the DAE controller has more knowledge and awareness of the system and can thus make the system more robust to disturbances as compared to the ODE-based controllers. Similar results have been achieved for Case $39$-bus system as presented in Figs. \ref{fig:case 39 delta-d 0.01 delta-I 0.3} and \ref{fig:case 39 worst scenior}. We can see that with the proposed robust WACs (in particular with $\mathcal{H}_{\infty}$-DAE) the power system is more robust and can withstand higher disturbances without losing synchrony.  
	
	\subsection{Updated Controllers under Norm-Bounded Parametric Uncertainty}
	
	It is noteworthy that all the proposed robust controllers have been designed subject to the assumption that we have access to an exact or highly-accurate model of the DAE-modeled power system. However, it is a strong assumption in some realistic cases in which there could exist a modeling error or parametric uncertainty on the state matrix $\m A$. To deal with such an issue, we propose the following two-step procedure:
	\begin{enumerate}
		\item First, we search for the worst-case scenario norm-bounded parametric uncertainty $ \m \Delta \m A$ given the information of the upper bound on its Frobenius norm (i.e., $\rho$ is given such that $\| \m \Delta \m A\|_F \le \rho$ holds).
		\item Second, we construct $\m A + \m \Delta \m A$ and \textit{update} the controller via redesigning it for the perturbed $\m A$, i.e., $\m A + \m \Delta \m A$. 
	\end{enumerate}
	To solve for the worst-case scenario norm-bounded parametric uncertainty $\m \Delta \m A$, we define the following function to be maximized:
	\begin{align}
		h(\m \Delta \m A) &:= \|\m G(s;\m \Delta \m A)\|_{\mathcal{H}_{\infty}} + \nu \alpha(\m A + \m \Delta \m A)
	\end{align}
	where $\|\m G(s;\m \Delta \m A)\|_{\mathcal{H}_{\infty}}$ denotes the $\mathcal{H}_{\infty}$ norm associated with the perturbed $\m A$ without any controller, $\alpha(\m A +\m \Delta \m A)$ represents the spectral abscissa (i.e., the maximum real part of the eigenvalues) of $\m A + \m \Delta \m A$, and $\nu$ is a penalizing parameter.
	Let us construct the following optimization problem:
	\begin{align*}
		\mathbf{\left(OP_4\right) }\;\;\;\;\;\minimize_{{\m \Delta A}} &\;\;\; -h(\m \Delta \m A) \\ \subjectto & \;\;\; \| \m \Delta \m A \|_F \le \rho. \;
	\end{align*}
	Solving $\mathbf{OP_4}$ for $\m \Delta \m A$, we obtain the worst-case scenario norm-bounded parametric uncertainty $\m \Delta \m A$ which ideally makes $\m A + \m \Delta \m A$ unstable with $\|\m G(s;\m \Delta \m A)\|_{\mathcal{H}_{\infty}} = \infty$. Note that since matrix $\m A$ intrinsically has a sparsity structure, we extract and impose such a sparsity structure to $\m \Delta A$ in parameterizing $\m \Delta A$ to feed into $\mathbf{OP_4}$.
	
	Setting $\rho = 1$ and $\nu = 0.001$, we solve $\mathbf{OP_4}$ for $\m \Delta \m A$ via MATLAB built-in function \texttt{fmincon()} and use it to redesign the updated $\mathcal{H}_{\infty}$-DAE, $\mathcal{H}_{\infty}$-ODE, and $\mathcal{H}_{2}$-ODE controllers. Considering the $9$-bus test system, Table \ref{table:1} shows the $\mathcal{H}_{\infty}$/$\mathcal{H}_2$ optimal values for three different cases: \textit{(i)} nominal controller, \textit{(ii)} updated controller under the random parametric uncertainty, and \textit{(iii)} updated controller under the worst-case scenario parametric uncertainty.
	{\small \begin{table}[t]
			\caption{ The values of $\|G_{cl}(s) \|_{\mathcal{H}_{\infty}}$/$\|G_{cl}(s) \|_{\mathcal{H}_{2}}$ for three different scenarios: \textit{(i)} nominal controller, \textit{(ii)} updated controller under the random parametric uncertainty, and \textit{(iii)} updated controller under the worst-case scenario parametric uncertainty.}
			\label{table:1}
			\centering
			\begin{tabular}{c|c|c|c} 
				\hline
				& $\mathcal{H}_{\infty}$-DAE & $\mathcal{H}_{\infty}$-ODE & $\mathcal{H}_{2}$-ODE \\ 
				\hline
				Nominal & $\infty$ & $\infty$ & $\infty$ \\\hline 
				Updated-R & $11.3678$ & $\infty$ & $\infty$ \\\hline
				Updated-WCS & $3.3546$ & $3.3547$ & $56.8720$\\
				\hline
			\end{tabular}
	\end{table}}
	As Table \ref{table:1} reflects, the nominal controller designs are not robust against the worst-case scenario parametric uncertainty. The updated controllers based on the random parametric uncertainty are not robust against the worst-case scenario parametric uncertainty except for the updated $\mathcal{H}_{\infty}$-DAE. Again, such an observation certifies that $\mathcal{H}_{\infty}$-DAE is the best approach among the proposed approaches. The updated controllers, taking advantage of the worst-case scenario $\m \Delta A$ information, are all robust against the worst-case scenario parametric uncertainty. In other words, they can successfully make the perturbed closed-loop system stable. Also, as shown by the column corresponding to $\mathcal{H}_{\infty}$-DAE in Table \ref{table:1}, the updated $\mathcal{H}_{\infty}$-DAE attains the best $\mathcal{H}_{\infty}$ performance compared to the nominal and the updated-R controllers. 
	
	
	
	\section{Paper Summary and Limitations}
	Given the detailed numerical simulations studies in the previous section, we make the following observations and thus answer the questions we posed in Section \ref{sec:case studies}:
	
	\begin{itemize}
		\item [--] \textit{A1:} With the proposed robust WAC as a secondary control loop the system frequency oscillations, RoCoF, and frequency nadir can be improved thus the overall system transient stability and dynamic response after a large disturbance can be made better. 
		\item [--] \textit{A2:} The designed NDAE-based feedback controller can add more damping to the system oscillations and can improve frequency nadir more as compared to the proposed NODE-based controllers. This shows that the more knowledge the feedback controller has about the system states (both dynamic and algebraic) the better control signal it can send to the power plants.  
		\item [--] \textit{A3:} The NDAE-based feedback controller is more robust toward disturbances and can keep the system synchronized for much higher disturbances in loads and renewables as compared to the NODE-based ones. Also, the $\mathcal{H}_\infty$-based controllers outperform $\mathcal{H}_2$-based controller, this corroborates with control theory's first principles.
		\item [--] \textit{A4:} Computing controller gain matrix for NDAE-based robust WAC can be much more challenging for a larger power network, while NODE-based feedback controller can easily be scaled to a larger power model. This shows the trade-off between computational tractability and performance---see A$2$ and A$3$ above.
	\end{itemize}
	\textcolor{black}{Next, we outline the paper's limitations and future work.
		First, in this work, we designed a time-invariant feedback controller. However, due to model mismatches and abnormalities dynamic feedback controller maybe be a better option at the cost of complexity and scalability. Second, in this work, we did not take into account, transmission delays or cyber-attacks in the communication network as it requires a new theoretical treatment. Lastly, the proposed control architecture is not sparse and requires all generators to participate in the control action which in turn requires a good and reliable communication network among all the power plants.} 
	Future work would be about addressing the above limitations, implementation of the proposed methodologies on various much bigger power systems, and comparing it with other advanced data driven nonlinear DAE controllers designs.
	
		

	
	\section*{Acknowledgements}
	The authors would like to acknowledge the insightful comments from the reviewers and the editor, who specifically contributed to improving the case studies section. We also acknowledge the support from the National Science Foundation (EPCN and DCSD programs) through Grants 2152450 and 2151571. Without NSF's support, this effort would not have been possible. 
	\bibliographystyle{unsrt}
	\bibliography{mybibfile}
	
	\newpage 
	
	\appendices
	
	\section{Traditional Power Plant Dynamics}\label{appndix:ninth Gen_dynamics}
	We consider a comprehensive $9^{th}$-order transient model of synchronous generator $i \in$ $\mathcal{G}$, which includes hydro/steam turbine, governor, and IEEE type DC1 excitation system dynamics given as follows \cite{sauer2017power}:
	\begin{itemize}[leftmargin=*]
		\begin{subequations} \label{eq:SynGen9th}
			\item Swing equations:	
			\begin{align}
				\begin{split}
					\hspace{-0.25cm}\dot{\delta}_{\mr{g}_i} &= \omega_{\mr{g}_i} - \omega_{0}\\ 
					\begin{split}
						\dot{\omega}_{\mr{g}_i} &=\dfrac{1}{2H_i}( T_{\mr{M}_i}\hspace{-0.05cm}-\hspace{-0.05cm}T_{\mr{e}_i})\;\;\;\mr{with}\;\; T_{\mr{e}_i}\hspace{-0.08cm} =\hspace{-0.08cm} {E}_{\mr d_i}i_{\mr d_i} + {E}_{\mr q_i}i_{\mr q_i} \end{split}\\ 
					\hspace{-0.25cm}\dot{E}_{\mr q_i} &= -\frac{1}{t_{\mr{qo}_i}}(E_{\mr q_i}-(x'_{\mr{q}_i}-x_{\mr{q}_i})i_{\mr d_i})\\
					\dot{E}_{\mr d_i} &= -\frac{1}{t_{\mr{do}_i}}(E_{\mr d_i}+(x'_{\mr{d}_i}-x_{\mr{d}_i})i_{\mr q_i} - E_{\mr{fd}_i}). 			
				\end{split}
			\end{align}
			\item Turbine and governor dynamics:	
			\begin{align}\label{eq:gen_tur/gov-dyn}
				\begin{split}
					\hspace{-0.25cm}\dot{T}_{\mr{M}_i} &=
					\left\{
					\begin{array}{ll}
						-\frac{1}{t_{\mr{ch}i}} (T_{\mr{M}_i}-P_{v_i}) & \mr{if \;\; thermal}\\
						-\frac{2}{t_{wi}} (T_{\mr{M}_i}-P_{v_i}+ t_{\mr{ch}i}\dot P_{v_i})& \mr{if \;\; hydro}
					\end{array}
					\right.\\
					\dot{P}_{v_i} &= -\frac{1}{t_{vi}}(P_{v_i}-P^*_{v_i} + \dfrac{\omega_i - 1}{R_{di}}).
				\end{split}
			\end{align}
			\item Excitation system dynamics:	
			\begin{align}\label{eq:gen_excit_dyn}
				\begin{split}
					\begin{split}
						\hspace{-0.25cm}\dot{E}_{\mr{fd}_i} \hspace{-0.05cm}&=\hspace{-0.05cm} \frac{-1}{t_{\mr{fd}i}}(k_{ei}+S_{ei}E_{\mr{fd}_i} - v_{ai})\;\;\mr{with}\;\; S_{ei}\hspace{-0.05cm} =\hspace{-0.05cm} a_i\mr{e}^{b_iE_{\mr{fd}i}}\end{split}\\
					\dot{r}_{f_i} &= -\frac{1}{t_{fi}}(r_{f_i} - \dfrac{k_{fi}}{t_{fi}}E_{\mr{fd}_i})\\
					\dot{v}_{ai} &= -\frac{1}{t_{ai}}(v_{ai} -k_{ai}v_{ei})
				\end{split}
			\end{align} 
		\end{subequations}
	\end{itemize}
	where $\delta_{\mr{g}_i}$ is the generator rotor angle $\mr{(pu)}$, $ \omega_{\mr{g}_i}$ is the generator rotor speed $\mr{(pu)}$, $\omega_0$ denotes synchronous speed $\mr{(pu)}$, $E_{\mr q_i}$, $E_{\mr d_i}$ are the dq-axis voltages $\mr{(pu)}$ behind transient reactance, $x'_{\mr q_i}, x'_{\mr d_i}, x_{\mr q_i},x_{\mr d_i}$ are the generator transient reactance and reactance $\mr{(pu)}$ along dq-axis, $i_{\mr d_i}$, $i_{\mr q_i}$ are the dq-axis generator currents,  $t_{\mr{qo}_i}$, $t_{\mr{do}_i}$ are the open circuit time constants ($\mr{sec}$) along dq-axis,  ${T}_{\mr{M}_i} $ represents mechanical torque $\mr{(pu)}$, ${T}_{\mr{e}_i}$ is the electrical air gap torque $\mr{(pu)}$, $P_{v_i}$ denotes steam/hydro valve position $\mr{(pu)}$, $P^*_{v_i}$ is the valve position set point $\mr{(pu)}$ from the grid operator, $E_{\mr{fd}_i}$ denotes generator field voltage $\mr{(pu)}$, $v_{ai}$ represents amplifier voltage $\mr{(pu)}$, $r_{f_i}$ is the stabilizer output $\mr{(pu)}$, $H_i$ is the generator inertia constant ($\mr{pu} \times \mr{sec}$), $R_{di}$ is the governor droop constant $(\mr{Hz}/\mr{pu})$, and  $k_{ei}$, $k_{fi}$, $k_{ai}$ are the exciter, stabilizer, and amplifier gains $\mr{(pu)}$, respectively.
	
	Moreover, in \eqref{eq:SynGen9th}, $t_{\mr{ch}i}$, $t_{wi}$, $t_{vi}$, $t_{\mr{fd}i}$, $t_{fi}$, and $ t_{ai}$ are the time constants ($\mr{sec}$) for steam/hydro valve position, field voltage, stabilizer, and amplifier respectively while $S_{ei}$ represents generator field voltage saturation function with constant scalars $a_i$, $b_i$ as detailed in \cite{sauer2017power}. Similarly, $v_{ei}$ in \eqref{eq:gen_excit_dyn} is the voltage control error and is given as: $v_{ei} = V_{i}^* -V_i+ r_f - \dfrac{k_{fi}}{t_f}E_{\mr{fd}_i}$ where $V_{i}^*$ is the voltage set point from the grid operator and  $V_i$ is the generator terminal voltage. 
	
	\section{Grid-Forming PV Plant Dynamics}\label{appndix:Grid forming PV dynamics}
	The dynamic equations describing all the states of solar power plants used in this study are detailed as follows \cite{SoumyaITPWRS2022, WasynczukITPE1996}:
	\begin{itemize}[leftmargin=*]
		\begin{subequations} \label{eq:PVdyn}
			\item DC side dynamics:
			\begin{align}\label{eq:DC_link_dyn}
				\dot{E}_{\mr{dc}_i} \hspace{-0.08cm}=\hspace{-0.08cm} \dfrac{1}{B_{C_i}}\left( P_{\mr{pv}_i}\hspace{-0.08cm} -\hspace{-0.08cm} P_{{\mr c}_i} \right).
			\end{align}
			\item AC side dynamics:
			\begin{align}\label{eq:AC_dyn}
				\begin{split}
					\hspace{-0.2cm}\dot i_{\mr{df}_i} &= \dfrac{\omega_b}{X_{f_i}}\left( -r_{f_i}i_{\mr{df}_i}\hspace{-0.1cm}+\hspace{-0.051cm} \omega_{c_i}X_{f_i}i_{\mr{qf}_i} \hspace{-0.051cm}+\hspace{-0.051cm} v_{\mr{df}_i}- v_{\mr{do}_i}\right) \\
					\hspace{-0.2cm}\dot i_{\mr{qf}_i} &= \dfrac{\omega_b}{X_{f_i}}\left( -r_{f_i}i_{\mr{qf}_i}\hspace{-0.1cm}+\hspace{-0.051cm} \omega_{c_i}X_{f_i}i_{\mr{df}_i} \hspace{-0.051cm}+\hspace{-0.051cm} v_{\mr{qf}_i}- v_{\mr{qo}_i}\right)\\
					\hspace{-0.2cm}\dot v_{\mr{dc}_i} &= \dfrac{\omega_b}{B_{c_i}}\left( \omega_{c_i}B_{c_i}v_{\mr{qc}_i} + i_{\mr{df}_i} - i_{\mr{dg}_i}\right) \\
					\hspace{-0.2cm}\dot v_{\mr{qc}_i} &= \dfrac{\omega_b}{B_{c_i}}\left( \omega_{c_i}B_{c_i}v_{\mr{dc}_i} + i_{\mr{qf}_i} - i_{\mr{qg}_i}\right) \\
					\hspace{-0.2cm}\dot\delta_{c_i} \hspace{-0.01cm}&=\hspace{-0.01cm} \omega_b(\omega_{c_i} \hspace{-0.01cm}- \hspace{-0.01cm}\omega_0) \;\; \mr{with}\;\; \omega_{c_i} \hspace{-0.08cm}=\hspace{-0.08cm} 1\hspace{-0.04cm}-\hspace{-0.04cm}k_{p_i}({\tilde{P}}_{e_i}\hspace{-0.04cm}-\hspace{-0.04cm}P^*_{e_i})\\
					\hspace{-0.2cm}\dot{\tilde{P}}_{e_i}\hspace{-0.08cm} &=\hspace{-0.08cm} \dfrac{1}{\tau_{s_i}}(-\tilde{P}_{e_i} \hspace{-0.04cm}+\hspace{-0.04cm} {P}_{e_i} )\\
					\hspace{-0.2cm}\dot{\tilde{Q}}_{e_i} \hspace{-0.04cm}&=\hspace{-0.04cm} \dfrac{1}{\tau_{s_i}}(-\tilde{Q}_{e_i}\hspace{-0.08cm} + \hspace{-0.08cm}{Q}_{e_i} ).
				\end{split}
			\end{align}
			\item Voltage regulator dynamics:
			\begin{align}\label{eq:vol_reg}
				\begin{split}
					\dot z_{\mr{do}_i}\hspace{-0.08cm} &=\hspace{-0.08cm} \dfrac{\kappa_{\mr{pv}_i}}{\tau_{v_i}}(v^*_{\mr{do}_i}\hspace{-0.08cm}-\hspace{-0.08cm}v_{\mr{do}_i})\;\;\;\mr{with}\;\;\;v^*_{\mr{do}_i}\hspace{-0.04cm}=\hspace{-0.04cm} V^*_i\hspace{-0.04cm}+\hspace{-0.04cm}k_{d_i}i_{\mr{qg}_i}\\
					\dot z_{\mr{qo}_i}\hspace{-0.08cm} &=\hspace{-0.08cm} \dfrac{\kappa_{\mr{pv}_i}}{\tau_{v_i}}(v^*_{\mr{qo}_i}\hspace{-0.08cm}-\hspace{-0.08cm}v_{\mr{qo}_i})\;\;\;\mr{with}\;\;\;v^*_{\mr{qo}_i}\hspace{-0.1cm}=0.
				\end{split}
			\end{align}	 
			\item Current regulator dynamics:
			\begin{align}\label{eq:Currt_reg_dyn}
				\begin{split}
					\dot z_{\mr{df}_i}\hspace{-0.01cm} &=\hspace{-0.01cm} \dfrac{\kappa_{\mr{p}_i}}{\tau_{i_i}}(i^*_{\mr{df}_i}\hspace{-0.01cm}-\hspace{-0.01cm}i_{\mr{df}_i})\\
					i^*_{\mr{df}_i}\hspace{-0.01cm}&=\hspace{-0.01cm} \hspace{-0.01cm}\kappa_{\mr{pv}_i}(v^*_{\mr{do}_i}-v_{\mr{do}_i}+ z_{\mr{do}_i}+i_{\mr{dg}_i}+i_{\mr{dc}_i})\\
					\dot z_{\mr{qf}_i}\hspace{-0.01cm} &=\hspace{-0.01cm} \dfrac{\kappa_{\mr{p}_i}}{\tau_{i_i}}(i^*_{\mr{qf}_i}\hspace{-0.01cm}-\hspace{-0.01cm}i_{\mr{qf}_i})\\
					i^*_{\mr{qf}_i}\hspace{-0.01cm}&=\hspace{-0.01cm} \hspace{-0.01cm}\kappa_{\mr{pv}_i}(v^*_{\mr{qo}_i}-v_{\mr{qo}_i}+ z_{\mr{qo}_i}+i_{\mr{qg}_i}+i_{\mr{qc}_i}).
				\end{split}
			\end{align}
		\end{subequations}
	\end{itemize}
	
	In \eqref{eq:DC_link_dyn}, $E_{\mr{dc}_i}$ is the energy stored in the DC link capacitor, $B_{C_i}$ is the DC link capacitance, $P_{\mr{pv}_i}$ is the DC power supplied by the PV array, and $P_{{\mr c}_i}$ is the power withdrawn by the GFM inverter. 
	Similarly, in \eqref{eq:AC_dyn}, $r_{f_i}$, $X_{f_i}$ are the resistance and reactance of LCL filter, $r_{c_i}$, $B_{c_i}$ are the resistance and capacitance of the capacitor in the LCL filter, $i_{\mr{df}_i}$, $i_{\mr{qf}_i}$ are the dq-axis current flowing through the LCL filter, $i_{\mr{dc}_i}$, $i_{\mr{qc}_i}$ are the dq-axis current flowing in the capacitor of LCL filter, $i_{\mr{dg}_i}$, $i_{\mr{qg}_i}$ are the dq-axis current flowing to the grid, $v_{\mr{dc}_i}$, $v_{\mr{qc}_i}$ are dq-axis capacitor voltage of the LCL filter, $\omega_b$ is the base speed ($\mr{pu}$), $\omega_{c_i}$ is the inverter angular speed ($\mr{pu}$), $k_{p_i}$ is the inverter droop constant, $\tau_{s_i}$ is the time constant of the low pass filter, $P_{e_i}$, $Q_{e_i}$ are the active and reactive power flowing to the grid ($\mr{pu}$), $P^*_{e_i}$ is the active power set-point commanded by the grid operator, while $\tilde{P}_{e_i}$,  $\tilde{Q}_{e_i}$ are the phasor representations of ${P}_{e_i}$,  ${Q}_{e_i}$ after passing through low pass filter (which are later used in the droop control of the GFM inverter). 
	
	Furthermore, in \eqref{eq:vol_reg} and \eqref{eq:Currt_reg_dyn}, $k_{q_i}$ is the voltage droop constant, $V^*_i$ is the voltage set point command by the grid operator while $\tau_{v_i}$, $\tau_{i_i}$, $\kappa_{\mr{pv}_i}$, $\kappa_{\mr{p}_i}$ are the time constants and gains of voltage and current regulator, respectively. Notice that voltage and current regulation in the GFM inverter is simply achieved by a proportional-integral (PI) type controller with $z_{do}$, $z_{qo}$, $z_{df}$, $z_{qf}$ representing the dq-axis states of integral compensator \cite{WasynczukITPE1996}.
	\section{Proof of Theorem \ref{theorm:H_inf}}\label{appndix:Proof therm1}
	Before we begin the proof, for the sake of simplicity in designing a controller gain in a tractable way, let us assume that the perturbation in nonlinear function in the closed-loop dynamics \eqref{eq:final_NDAE_peturbed_final} is $\mathcal{L}_2$-norm bounded and can be written as $\Delta\m f(\m x,\m u_{cl},\m w) = \mB_f\m w_f$, where $\m B_f = \mB_w$. Notice that this assumption is just carried out to simplify the matrix inequalities in the controller design, the overall controller in the end is applied to the whole NDAE power system model \eqref{eq:final_NDAE_cntrl} without any loss of generality. We now define:
	\begin{align}\label{eq:B_w tilde}
		\tilde{\m w}\hspace{-0.02cm} =\hspace{-0.02cm} \bmat{\m w^\top & \m w_f^\top}^\top,~ \hat{\mB}_w \hspace{-0.02cm}=\hspace{-0.02cm} \bmat{\mB_w & \m B_f},~\hat{\mD}_w\hspace{-0.02cm} = \hspace{-0.02cm}\bmat{\mD_w & \m D_f}
	\end{align}
	where $\mD_f = \mB_f$. Substituting \eqref{eq:B_w tilde} in \eqref{eq:final_NDAE_peturbed_final}, we obtain the following equivalent perturbed closed-loop dynamics:
	\begin{subequations}\label{eq:Hinf_NDAE_peturbed}
		\begin{align}
			\m E\dot{\m x} &= (\m A+\m{BK})\m x+ \hat{\m B}_w \tilde{\m w}\\
			\m z_1 &= (\m C+\m{DK})\m x + \hat{\mD}_w\tilde{\m w}.
		\end{align}
	\end{subequations}
	That being said, the proof of Theorem \ref{theorm:H_inf} is given as follows:
	Let us consider a quadratic Lyapunov candidate function $V(\m x)=\m x^\top \mE^\top \mP \m x $, where $V:\mbb{R}^{n}\rightarrow \mbb{R}_+$, $\m P\in\mbb{R}^{n\times n}$, and $\mE^\top \mP = \mP^\top \mE \succeq \mO$, then its derivative along trajectories of $\m x$ can be written as
	\begin{align*}
		{\dot V}(\m x) &= (\m{E}\dot{\m x})^\top \mP \m x+(\m{E}{\m x})^\top (\mP \dot{\m x}).
	\end{align*}
	Since $\mE^\top \mP = \mP^\top \mE $, then we can write
	\begin{align*}
		{\dot V}(\m x) &= (\m{E}\dot{\m x})^\top \mP \m x+(\mP \m x)^\top(\m{E}\dot{\m x}).
	\end{align*}
	Substituting value of $\m{E}\dot{\m x}$ from \eqref{eq:Hinf_NDAE_peturbed} yields
	\begin{align*}
		{\dot V}(\m x) = ( \m{A}_{cl}\m x + \hat{\m B}_w \tilde{\m w})^\top\m{Px} + {(\m P\m x)}^\top( \m{A}_{cl}\m x + \hat{\m B}_w \tilde{\m w})
	\end{align*}
	where $ \m{A}_{cl} = \m A+\mB\mK$. From Def. \ref{def:H_inf} for any $\mathcal{L}_2$-norm bounded uncertainty $\tilde{\m w}$, the $\mathcal{H}_{\infty}$ stability criterion with Lyapunov stability can be written as ${\dot V}(\m x) + \m z_1^\top\m z_1 - \mu^2\tilde{\m w}^\top\tilde{\m w} < 0$, i.e., 
	\begin{align*}
		&( \m{A}_{cl}\m x + \hat{\m B}_w \tilde{\m w})^\top\m{Px} + {(\m P\m x)}^\top( \m{A}_{cl}\m x + \hat{\m B}_w \tilde{\m w})+\\&\m x^\top(\mC+\m{DK})^\top(\mC+\m{DK})\m x + \tilde{\m w}^\top\hat{\m D}_w^\top\hat{\m D}_w\tilde{\m w}-\\&\mu^2\tilde{\m w}^\top\tilde{\m w}+
		\m x^\top(\mC+\m{DK})^\top\hat{\m D}_w\tilde{\m w} + \tilde{\m w}^\top\hat{\m D}_w^\top(\mC+\m{DK})<0. 
	\end{align*}
	The above equation can be rewritten as $\m\Phi^\top\m\Xi\m\Phi<0$ where
	\begin{align*}
		\m\Phi\hspace{-0.1cm} = \hspace{-0.1cm}\bmat{\m x\\ \m w}, \m\Xi=\hspace{-0.1cm}\bmat{ \mA^\top_{cl}\mP+\mP^\top \mA_{cl}+\m\Gamma^\top\m\Gamma& \mP^\top\hat{\m B}_w + \m\Gamma^\top\hat{\m D}_w \\ \hat{\m B}_w^\top\mP + \hat{\m D}_w^\top\m\Gamma& \hat{\m D}_w^\top\hat{\m D}_w-\mu^2\mI}
	\end{align*}
	with $\m\Gamma =\mC+\m{DK}$. Note that $\m\Phi^\top\m\Xi\m\Phi<0$ holds only if $\m\Xi\prec \mO$. Now, assuming $\mP$ is invertible and defining $\mS=\mP^{-1}$, then applying congruence transformation to $\m\Xi$ as:
	\begin{align*}
		\bmat{\mS^\top\mO\\ \mO\;\;\;\;\mI} \bmat{ \mA^\top_{cl}\mP\hspace{-0.05cm}+\hspace{-0.05cm}\mP^\top \mA_{cl}\hspace{-0.05cm}+\hspace{-0.05cm}\m\Gamma^\top\m\Gamma& \mP^\top\hat{\m B}_w \hspace{-0.05cm}+\hspace{-0.05cm} \m\Gamma^\top\hat{\m D}_w \\ \hat{\m B}_w^\top\mP \hspace{-0.05cm}+\hspace{-0.05cm} \hat{\m D}_w^\top\m\Gamma& \hat{\m D}_w^\top\hat{\m D}_w\hspace{-0.05cm}-\hspace{-0.05cm}\mu^2\mI} \bmat{\mS\;\;\mO\\ \mO\;\;\mI} 
	\end{align*}
	which can be simplified to:
	\begin{align}\label{eq:prof_21}
		\hspace{-0.1cm}\bmat{ \mS^\top\mA^\top_{cl}\hspace{-0.05cm}+\hspace{-0.05cm} \mA_{cl}\mS\hspace{-0.05cm}+\hspace{-0.05cm}\mS^\top\m\Gamma^\top\m\Gamma\mS& \hat{\m B}_w+ \mS^\top\m\Gamma^\top\hat{\m D}_w \\ \hat{\m B}_w^\top + \hat{\m D}_w^\top\m\Gamma\mS& \hat{\m D}_w^\top\hat{\m D}_w\hspace{-0.05cm}-\hspace{-0.05cm}\mu^2\mI}\prec \mO.
	\end{align}
	Applying the Schur complement lemma \cite{zhang2006schur} then \eqref{eq:prof_21} can equivalently be rewritten as follows:
	\begin{align}\label{eq:proof_BMI}
		\bmat{ \mS^\top\mA^\top_{cl}+\mA_{cl}\mS &\hat{\m B}_w& \mS^\top\m\Gamma^\top \\ 
			\hat{\m B}_w^\top & -\mu^2\mI & \hat{\m D}_w^\top\\
			\m\Gamma\mS & \hat{\m D}_w &-\mI}\prec \mO.
	\end{align}
	Now, notice that at the beginning of the proof we assumed $\mE^\top \mP = \mP^\top \mE \succeq \mO$, since this assumption has equality term, we need to remove this condition to get a strict LMI-based controller design. This can be achieved by finding the value of $\mS$ (which is equal to $\mP^{-1}$) from  $\mE^\top \mP = \mP^\top \mE$ and substituting it in \eqref{eq:proof_BMI}.
	To do so, let us assume there exist matrices 
	${\m \varTheta}\in \mbb{R}^{n\times n}$ and ${\m \varOmega}\in \mbb{R}^{n\times n}$ such that 
	\begin{align}\label{eq:proof-eq-1}
		\begin{split}
			\m{\varTheta E \varOmega} &= \bmat{\m I & \m O\\ \mO & \mO}, 
		\end{split}
		\begin{split}
			{(\m\varTheta^{-1})^\top \mP \m\varOmega} &= \bmat{\mP_1 & \mP_2\\ \mP_3 & \mP_4} 
		\end{split}
	\end{align}
	where $\mP_1 \in  \mbb{R}^{n_d\times n_d}$, $\m P_2 \in  \mbb{R}^{n_d\times n_a}$, $\mP_3 \in  \mbb{R}^{n_a\times n_d}$ and $\mP_4 \in  \mbb{R}^{n_a\times n_a}$. Then from \eqref{eq:proof-eq-1} we get 
	\begin{subequations}\label{eq:proof-eq-2}
		\begin{align}
			\begin{split}
				\m{E} &= \m\varTheta^{-1}\bmat{\m I & \m O\\ \mO & \mO}\m\varOmega^{-1}
			\end{split}\\
			\begin{split}
				\m{P} &= \m\varTheta^\top\bmat{\mP_1 & \mP_2\\ \mP_3 & \mP_4}\m\varOmega^{-1}. 
			\end{split}
		\end{align}
	\end{subequations}
	Now, from \eqref{eq:proof-eq-2}, $\mE^\top \mP$ and $\mP^\top \mE$ can be written as:
	\begin{subequations}\label{eq:proof-eq-3}
		\begin{align}
			\begin{split}\label{eq:proof-eq-3a}
				\mE^\top \mP = (\m\varOmega^{-1})^\top\bmat{\mP_1&\mO\\\mO&\mO}\m\varOmega^{-1}
			\end{split}\\
			\begin{split}\label{eq:proof-eq-3b}
				\mP^\top \mE = (\m\varOmega^{-1})^\top\bmat{\mP_1^\top&\mO\\\mP_2^\top&\mO}\m\varOmega^{-1}.
			\end{split}
		\end{align}
	\end{subequations}
	We can see from \eqref{eq:proof-eq-3} that $\mE^\top \mP$ and $\mP^\top \mE$ can be made equal only if $\m{P_2}^\top = \mO$ and $\m{P_1} = \m{P_1}^\top$. Hence, the value of $\mP$ can be updated as
	\begin{align*}
		\begin{split}
			\m{P} &\hspace{-0.0cm}= \hspace{-0.0cm}\m\varTheta^\top\underbrace{\bmat{\mP_1 & \m {O}\\ \mP_3 & \mP_4}}_{\m{\bar P}}\m\varOmega^{-1}.
		\end{split}
	\end{align*}
	Then, $\mS = \mP^{-1}$ can be written as
	\begin{align*}
		\begin{split}
			\m{S} &\hspace{-0.0cm}= \hspace{-0.0cm}\m\varOmega\underbrace{\bmat{\mP^{'}_1 & \m {O}\\ \mP^{'}_3 & \mP^{'}_4}}_{\m{\bar P}^{-1}}(\m\varTheta^{-1})^\top
		\end{split}
	\end{align*}
	\begin{align*}
		\begin{split}
			&\hspace{-0.0cm}= \hspace{-0.01cm}\m\varOmega\left( \bmat{\mP^{'}_1\;\;\;\m {O}\\ \m{O} \;\;\;\;\;\, \m{I}}\bmat{\m {I} \;\;\;\;\, \m {O}\\ \m{O}\;\;\m{O}}\right) (\m\varTheta^{-1})^\top\hspace{-0.02cm} + \hspace{-0.02cm}\m\varOmega\bmat{\m {O} \;\;\;\;\;\m {O}\\ \mP^{'}_3\;\;\mP^{'}_4}(\m\varTheta^{-1})^\top.
		\end{split}
	\end{align*}
	Let us define $\mX$ as 
	\begin{align*}
		\m{X} &= \m\varOmega\bmat{\mP^{'}_1 & \m {O}\\ \m{O} & \m{I}}\m\varOmega^\top.
	\end{align*}
	It is then straightforward to compute $\mS$ from the above equation as:
	\begin{subequations}\label{eq:proof-eq-final}
		\begin{align}
			\begin{split}
				\m{S} &= \m{XE}^\top + \underbrace{\m\varOmega\bmat{\mO\\\mI}}_{\mE^{\perp}}\underbrace{\bmat{\mP^{'}_3&\mP^{'}_4}(\m\varTheta^{-1})^\top}_{\mW}
			\end{split}\label{eq:proof-eq-finala}\\
			\begin{split}
				\m{S} &= \m{XE}^\top+\mE^{\perp}{\mW}\label{eq:proof-eq-finalb}
			\end{split}
		\end{align}
	\end{subequations}
	where $\mW\in \mbb{R}^{n_a\times n}$ and $\mE^{\perp}\in\mbb{R}^{n\times n_a}$ is the orthogonal  complement of $\mE$. Finally, by defining $\mH = \m{KS}$, $\lambda=\mu^2$, and plugging the value of $\mS$ from \eqref{eq:proof-eq-finalb} into \eqref{eq:proof_BMI} we get the strict LMI \eqref{eq:LMI_Hinf}. This ends the proof. 
	
	\section{Derivation of $\mathbf{OP_2}$}\label{appndix:OP3}
	
	Let us consider a quadratic Lyapunov candidate function $V_d(\m x_d)=\m x_d^\top \m P_d \m x_d $, where $V:\mbb{R}^{n_d}\rightarrow \mbb{R}_+$, $\m P_d\in \mbb{R}^{n_d\times n_d}$, and $ \m P_d^\top = \m P_d \succ \mO$ hold. Then, its derivative along trajectories of $\m x_d$ can be written as
	\begin{align*}
		{\dot V}_d(\m x_d) &= \dot{\m x}_d^\top \m P_d \m x_d +{\m x}_d^\top \m P_d \dot{\m x}_d.
	\end{align*}
	Substituting the value of $\dot{\m x}_d$ from \eqref{eq:NODE_peturbed} yields
	\begin{align*}
		{\dot V}_d(\m x_d) = (\tilde{\m{A}}_{cl}\m x_d + \tilde{\m B}_w \tilde{\m w})^\top\m{P}_d \m x_d + {\m x}_d^\top \m P_d (\tilde{\m{A}}_{cl}\m x_d + \tilde{\m B}_w \tilde{\m w})
	\end{align*}
	where $ \tilde{\m{A}}_{cl} = \tilde{\m A}+\tilde{\mB}\m K_d$. From Def. \ref{def:H_inf} for any $\mathcal{L}_2$-norm bounded uncertainty $\tilde{\m w}$, the $\mathcal{H}_{\infty}$ stability criterion with Lyapunov stability can be written as ${\dot V}_d(\m x_d) + \m z_2^\top\m z_2 - \mu^2\tilde{\m w}^\top\tilde{\m w} < 0$, i.e., $\m\Phi_2^\top\m\Xi_2\m\Phi_2 < 0$ where
	\begin{align*}
		& \m\Phi_2 \hspace{-0.01cm} = \hspace{-0.01cm}\bmat{\m x_d\\ \tilde{\m w}},~ \m\Xi_2=\hspace{-0.1cm}\bmat{\tilde{\m{A}}_{cl}^\top \m P_d +\m P_d \tilde{\m{A}}_{cl}+ \tilde{\m \Gamma}^\top \tilde{\m \Gamma}& \mP_d \tilde{\m B}_w + \tilde{\m\Gamma}^\top\tilde{\m D}_w \\ \tilde{\m B}_w^\top\mP_d + \tilde{\m D}_w^\top \tilde{\m\Gamma}& \tilde{\m D}_w^\top \tilde{\m D}_w-\mu^2\mI}
	\end{align*}
	with $\tilde{\m\Gamma} = \tilde{\mC}+ \tilde{\m{D}} \m K_d$. Note that $\m\Phi_2^\top\m\Xi_2\m\Phi_2 < 0$ holds only if $\m\Xi_2 \prec \mO$. Applying the Schur complement lemma \cite{zhang2006schur} to $\m\Xi_2 \prec  \mO$, we get
	\begin{align*}
		\hspace{-0.5cm}&\tilde{\m{A}}_{cl}^\top \m P_d \hspace{-0.02cm} + \hspace{-0.02cm}\m P_d \tilde{\m{A}}_{cl}\hspace{-0.02cm}+ \hspace{-0.02cm}\tilde{\m \Gamma}^\top \tilde{\m \Gamma}\hspace{-0.02cm} +\hspace{-0.02cm} \notag \\& (\mP_d \tilde{\m B}_w \hspace{-0.02cm}+\hspace{-0.02cm} \tilde{\m\Gamma}^\top\tilde{\m D}_w) (\mu^2\mI\hspace{-0.02cm}-\hspace{-0.02cm}\tilde{\m D}_w^\top \tilde{\m D}_w)^{-1} (\tilde{\m B}_w^\top\mP_d \hspace{-0.02cm}+\hspace{-0.02cm} \tilde{\m D}_w^\top \tilde{\m\Gamma}) \prec \mO
	\end{align*}
	that can be rewritten as
	\begin{align} \label{MIY2}
		&\bar{\m{A}}^\top \m P_d +\m P_d \bar{\m{A}}+ \m Q + \m P_d \m G \m P_d + \m K_d^\top \m R \m K_d +\notag \\& \m K_d^\top (\bar{\m B}^\top \m P_d + \m S^\top) + (\m P_d \bar{\m B} + \m S) \m K_d  \prec  \mO
	\end{align}
	where
	\begin{align*}
		\Bar{\mA}(\mu) &= \tilde{\m A}+\Tilde{\mB}_w\mF(\mu)^{-1}\Tilde{\mD}_w^\top\tilde{\mC}\\
		\Bar{\mB}(\mu) &= \tilde{\m B}+\Tilde{\mB}_w\mF(\mu)^{-1}\Tilde{\mD}_w^\top\tilde{\mD}\\
		{\mQ}(\mu) &= \tilde{\mC}^\top(\mI+\Tilde{\mD}_w\mF(\mu)^{-1}\Tilde{\mD}_w^\top)\tilde{\mC}\\
		{\mR}(\mu) &= \tilde{\mD}^\top(\mI+\Tilde{\mD}_w\mF(\mu)^{-1}\Tilde{\mD}_w^\top)\tilde{\mD}\\
		{\mS}(\mu) &= \tilde{\mC}^\top(\mI+\Tilde{\mD}_w\mF(\mu)^{-1}\Tilde{\mD}_w^\top)\tilde{\mD}\\
		{\mG}(\mu) &= \Tilde{\mB}_w\mF(\mu)^{-1}\Tilde{\mB}_w^\top\\
		{\mF}(\mu) &= \mu^2\mI-\Tilde{\mD}_w^\top\Tilde{\mD}_w.
	\end{align*}
	
	Completing the square in \eqref{MIY2}, we obtain
	\begin{align*}
		&\bar{\m{A}}^\top \m P_d \hspace{-0.04cm}+\hspace{-0.04cm}\m P_d \bar{\m{A}}\hspace{-0.04cm} + \hspace{-0.04cm}\m P_d \mG \m P_d \hspace{-0.04cm}-\hspace{-0.04cm} (\m P_d \bar{\m B}\hspace{-0.04cm} +\hspace{-0.04cm} \m S) \m R^{-1} (\bar{\m B}^\top \m P_d\hspace{-0.04cm} + \hspace{-0.04cm}\m S^\top) \notag\hspace{-0.03cm}  +\hspace{-0.03cm}  \mQ + \\&  (\m K_d^\top\hspace{-0.04cm} + \hspace{-0.04cm}(\m P_d \bar{\m B}\hspace{-0.04cm} +\hspace{-0.04cm} \m S) \m  R^{-1}) \m R (\m K_d \hspace{-0.04cm}+\hspace{-0.04cm} \m R^{-1} (\bar{\m B}^\top \m P_d \hspace{-0.03cm}+\hspace{-0.03cm} \m S^\top)) \hspace{-0.01cm}\prec\hspace{-0.01cm} \mO.
	\end{align*}
	Now, since $(\m K_d^\top + (\m P_d \bar{\m B} + \m S) \m  R^{-1}) \m R (\m K_d + \m R^{-1} (\bar{\m B}^\top \m P_d + \m S^\top)) \succeq \mO$ holds, we conclude that $\m K_d = -\m R^{-1} (\bar{\m B}^\top \m P_d + \m S^\top)$ holds. Then, we compute the $\mathcal{H}_\infty$ state-feedback controller $\mK$ as
	\begin{align*}
		\mK(\mu) = \bmat{-\mR(\mu)^{-1}(\Bar{\mB}(\mu)^\top\mP_d(\mu)+\mS(\mu)^\top)&\mO}
	\end{align*}
	for which the $\mathcal{H}_\infty$ norm of the transfer function from $\tilde{\m w}$ to $\m z_2$
	is less than $\mu$, via solving the following CARE:
	\begin{align*}
		\hspace{-0cm}\Bar{\mA}^\top\hspace{-0.1cm}\mP_d \hspace{-0.02cm} + \hspace{-0.02cm}\mP_d\Bar{\mA}\hspace{-0.01cm}+\hspace{-0.01cm}\mP_d\mG\mP_d\hspace{-0.01cm}-\hspace{-0.01cm}(\mP_d\Bar{\mB}\hspace{-0.01cm}+\hspace{-0.01cm}\mS)\mR^{-1}(\Bar{\mB}^\top\hspace{-0.01cm}\mP_d\hspace{-0.01cm}+\hspace{-0.01cm}\mS^\top)\hspace{-0.01cm}+\hspace{-0.01cm}\mQ \hspace{-0.01cm}=\hspace{-0.01cm} \mO
	\end{align*}
	for $\m P_d$. This completes the derivation.

\end{document}